\documentclass[a4paper,fleqn,12pt]{elsarticle}

\usepackage[numbers]{natbib}
\usepackage[ruled]{algorithm2e}
\usepackage{subfigure}
\usepackage{booktabs}
\usepackage{multirow}
\usepackage{amsmath}
\usepackage{color}
\usepackage{amssymb}
\usepackage[left=3cm, right=3cm, lines=45, top=0.8in, bottom=1.0in]{geometry}
\usepackage{float}

\graphicspath{ {./} }

\newtheorem{definition}{Definition}
\newtheorem{problem}{Problem}

\begin{document}
	\begin{sloppypar}
		
		\begin{frontmatter}
			\let\WriteBookmarks\relax
			\def\floatpagepagefraction{1}
			\def\textpagefraction{.001}
			
			\title{Session-based Social and Dependency-aware Software Recommendation}                      
			
			\author[1]{Dengcheng Yan}
			\ead{yanzhou@ahu.edu.cn}
			
			\author[1]{Tianyi Tang}
			\ead{tangtianyi202012@163.com}
			
			\author[1]{Wenxin Xie}
			\ead{xiewxahu@foxmail.com}
			
			\author[1]{Yiwen Zhang\corref{mycorrespondingauthor}}
			\cortext[mycorrespondingauthor]{Corresponding author}
			\ead{zhangyiwen@ahu.edu.cn}
			
			\author[2]{Qiang He}
			\ead{qhe@swin.edu.au}
			
			\address[1]{School of Computer Science and Technology, Anhui University, Hefei 230601, China}
			
			\address[2]{School of Software and Electrical
				Engineering, Swinburne University of Technology, Melbourne 3122, Australia}
			
			\begin{abstract}
				With the increase of complexity of modern software, social collaborative coding and reuse of open source software packages become more and more popular, which thus greatly enhances the development efficiency and software quality. However, the explosive growth of open source software packages exposes developers to the challenge of information overload. While this can be addressed by conventional recommender systems, they usually do not consider particular constraints of social coding such as social influence among developers and dependency relations among software packages. In this paper, we aim to model the dynamic interests of developers with both social influence and dependency constraints, and propose the Session-based Social and Dependency-aware software Recommendation (SSDRec) model. This model integrates recurrent neural network (RNN) and graph attention network (GAT) into a unified framework. An RNN is employed to model the short-term dynamic interests of developers in each session and two GATs are utilized to capture social influence from friends and dependency constraints from dependent software packages, respectively. Extensive experiments are conducted on real-world datasets and the results demonstrate that our model significantly outperforms the competitive baselines.
			\end{abstract}

			\begin{keyword}
				Software recommendation \sep Social network \sep Dependency network \sep Graph neural network
			\end{keyword}

		\end{frontmatter}

\section{Introduction}
	
	Modern software is becoming more and more complicated and the development usually needs collaboration of a team and depends on a large number of third-party software packages, which promotes the wide adoption of social collaborative coding paradigm. Recently, social collaborative coding platforms such as GitHub have emerged to provide developers with abundant functionalities of social collaboration and technical development and produce a large amount of high-quality open source software packages. According to GitHub \footnote{https://github.com/about}, there are more than 56 million developers collaborating on more than 100 million software projects as of March, 2021.
	
	While the explosive growth of open source software packages will significantly fuel the prosperity of the software industry, it also exposes developers to the challenge of information overload. Developers often need to spend much time searching software packages they are interested in. To address this challenge, it is essential to introduce recommendation systems which have been proven powerful to deal with information overload problem in various fields \cite{rendle2012bpr, he2017neural,chen2016minimizing,zhang2021privacy,zhang2019covering}.
	
    Recently, conventional recommendation models have been applied to software recommendation \cite{ichii2009software, thung2013automated, he2020diversified}, but they usually do not consider either the dynamics of developers' interests \cite{jiang2020adapting} or particular constraints of social coding such as social influence among developers and dependency relations among software packages. 
    
\begin{figure}[htb]
		\centering
		\includegraphics[width=\linewidth]{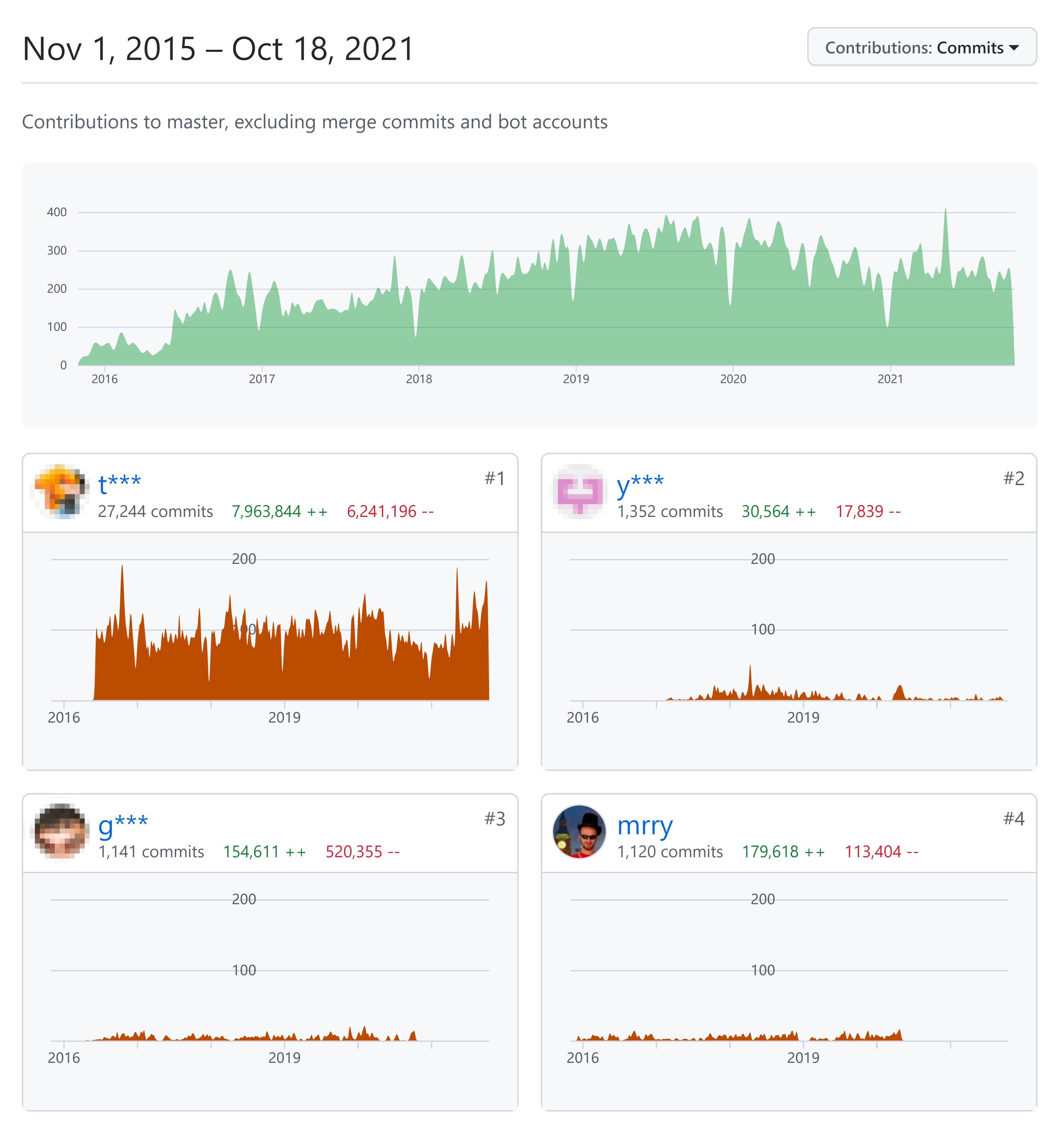}
		\caption{Contributors of project TensorFlow as of Oct. 18, 2021.\protect\footnotemark}
		\label{mrry-tf-contribution}
	\end{figure}

\begin{figure}[htb]
		\centering
		\includegraphics[width=\linewidth]{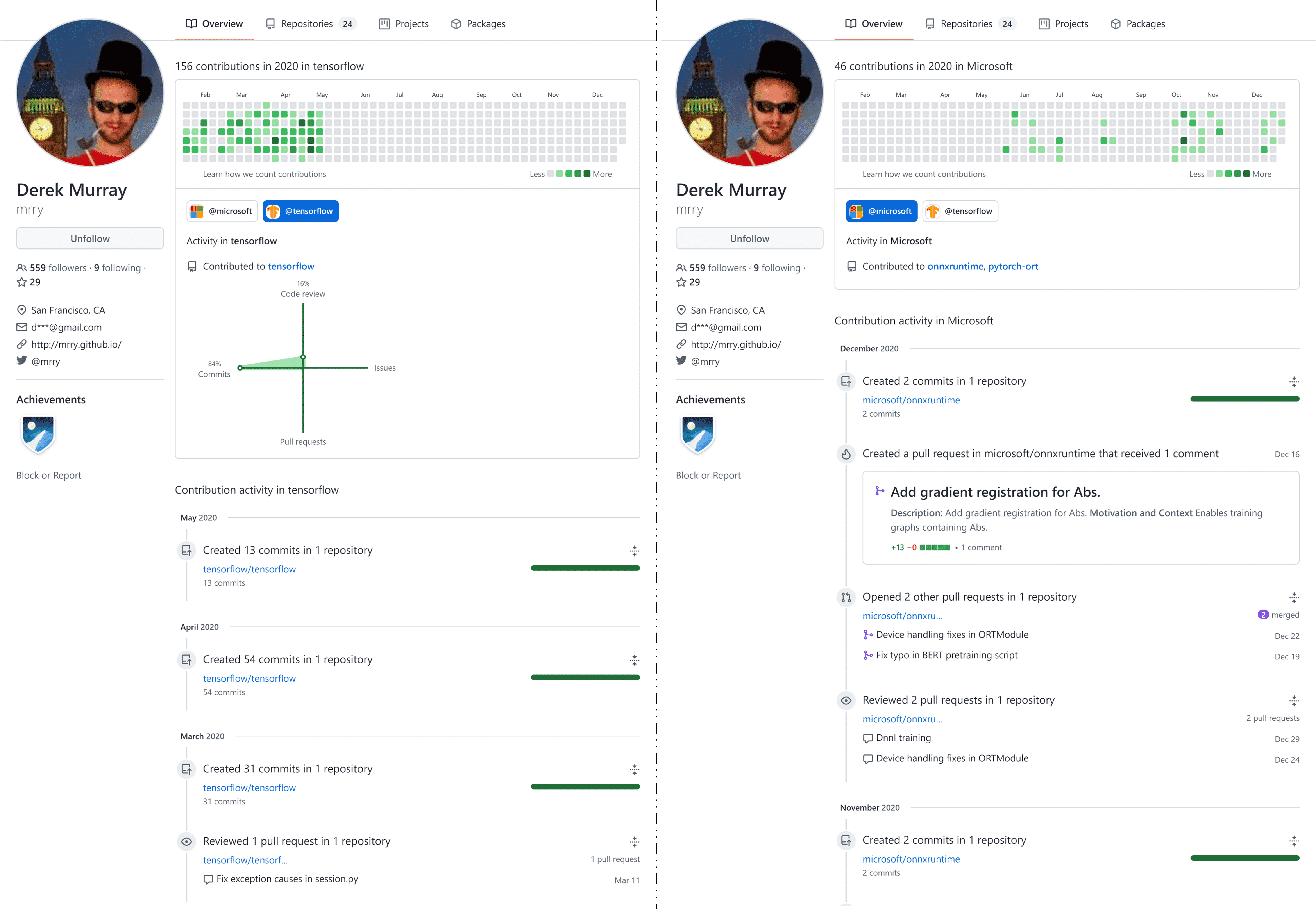}
		\caption{Dynamic interest of Derek Murray.\protect\footnotemark}
		\label{mrry-dynamic-interest}
	\end{figure}

    Actually, in the community of software development, developers’ interests are dynamic and gradually evolve due to the emergence of new project requirements or new techniques. At different time periods, a developer may focus on a different technical field and is interested in software packages related to that technical field. Taking Derek Murray (GitHub ID: mrry), a main contributor of the popular deep learning framework TensorFlow, as an example, we can find from Figure \ref{mrry-tf-contribution} that he began to contribute to TensorFlow upon its establishment in GitHub while his contribution stopped in the middle of the year 2020. From his profile page in GitHub (shown in Figure \ref{mrry-dynamic-interest}), we can find that in the middle of the year 2020 his interest evolves to ONNX, an open standard for machine learning interoperability, and its related projects. Thus, it is practical to capture developers’ dynamic interests in the scenario of software recommendation. Practically, the dynamic interest of each developer can be captured from the projects he/she recently starred or contributed by modeling the sequence of these projects with recurrent neural network (RNN). For example, for each month, we group the projects Derek Murray starred or contributed in time order and the embedding of each project (obtaining from the output of the GAT on dependency network) is then input to an RNN one by one. After all projects are finished, the final output of the RNN is obtained as the representation of Derek Murray’s interest in this time period.
    
	\footnotetext{Obtained from https://github.com/tensorflow/tensorflow/graphs/contributors  with privacy information hidden.}
    
	\footnotetext{Obtained from https://github.com/mrry with the consent of Derek Murray.}
    
\begin{figure}
		\centering
		\includegraphics[width=\linewidth]{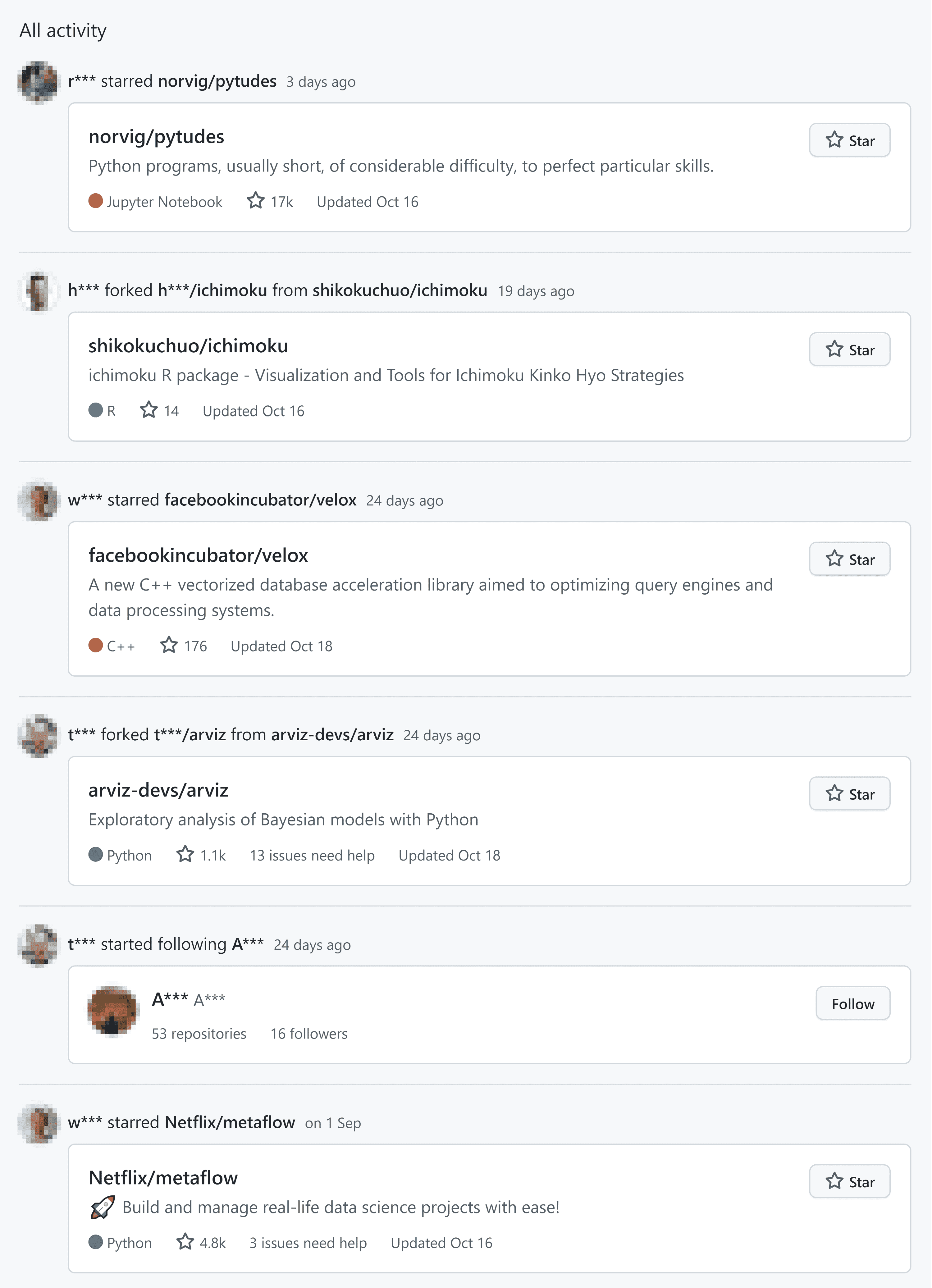}
		\caption{GitHub timeline of the author Dengcheng Yan.\protect\footnotemark}
		\label{github-timeline}
	\end{figure}

    Recently, as social collaborative coding is becoming a popular paradigm for software development, social collaborative coding platforms such as GitHub emerge and provide abundant social functionalities such as forming social relations with other developers in addition to plentiful software development tools. A developer in GitHub can follow other developers as friends and obtain in his/her timeline their recent behaviors/choices such as starring or creating a new repository. Then, the developer’s choice on software packages will be influenced by his/her friend developers’ choices because of social trust and time efficiency as there are large amount of software projects in GitHub, it is efficient to make choice based on friend developers’ choice. For example, Figure \ref{github-timeline} shows the GitHub timeline of one of the authors Dengcheng Yan. The recent behaviors of the developers Dengcheng Yan has followed appears in his GitHub timeline. As a developer, Dengcheng Yan frequently reads his GitHub timeline and he found developer r*** starred an interesting project norvig\/pytudes. Then he also starred this project. This phenomenon of social influence widely exists in GitHub and forms the intuition behind our proposal to employ social influence to enhance software recommendation. However, social relationship may have bad effect when a developer shift from one program language to another. So we should not treat all social relations equally. Actually, different social relations are weighted differently during different time periods. Specifically, we employ a graph attention network (GAT) on the social network and calculate the attention coefficient (i.e., strength of social influence) between developers by ingesting not only the social network structure but also the dynamic interest of developers. The more similar the dynamic interest between developers is, the stronger the social influence between them will be. In different time period, the dynamic interest of developers captures their recent preference and is different, thus making the strength of social influence adaptive to avoid outdated social relations.
    
	\footnotetext{Obtained from https://github.com with privacy information hidden.}
    
    Last and most importantly, software recommendation is different from conventional recommendation scenarios and dependency exists naturally between items (i.e., software packages) as a newly developed software package usually reuses the functionality of multiple existing software packages. Developers who develop the newly developed software packages should first be familiar with some of the functionalities of the dependency software packages. Thus, dependency constraints are another key factor that influences developers’ choices which is essential to be considered in the software recommendation scenario.
    
\begin{figure}[htb]
		\centering
		\includegraphics[width=\linewidth]{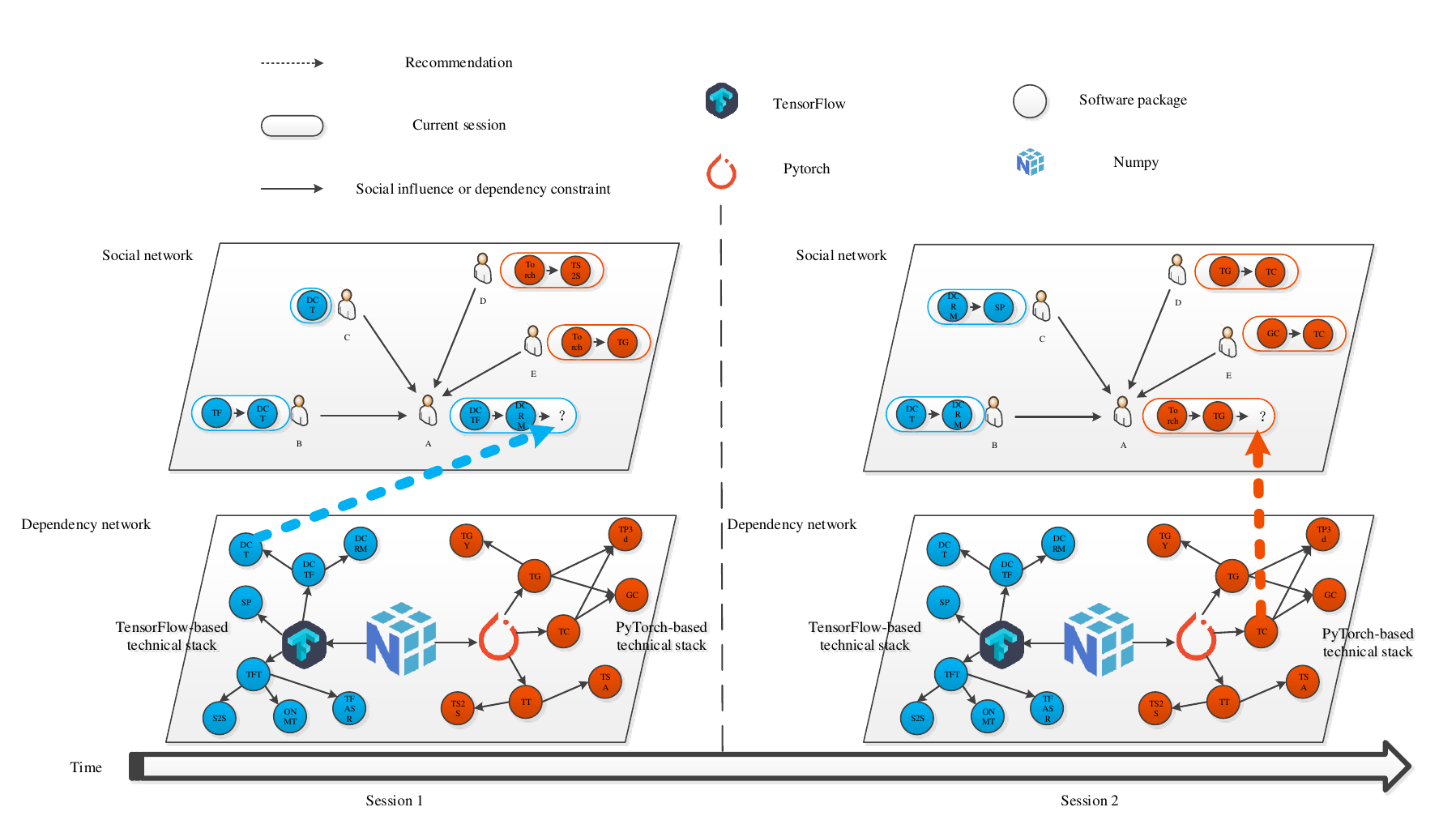}
		\caption{Motivation example. Due to space constraints, we utilize the abbreviations of the software packages to represent them and their full names are listed in table \ref{tab:Abbreviation}.}
		\label{motivation-example}
	\end{figure}

   We give a real-world motivation example which combines dynamic interest, social influence and dependency constraints of the popular python-based deep learning technical stacks, including both TensorFlow-based and PyTorch-based technical stacks(see Figure \ref{motivation-example}). In the first session, developer A has chosen DeepCell-tf (DCTF, for cell image analysis) and DeepCell-RetinaMask (DCRM, for cell object detection), we could conclude from both social influence and dependency constraints that he/she mainly focuses on developing tools for cell image analysis and would further extend his/her interest from simply detecting cells to tracking cells for cell lineage construction using DeepCell-Tracking (DCT). Then after some period, in the second session, developer A would further analyze the cell lineage he/she has constructed but finds no suitable tools in TensorFlow-based technical stack. He/she turns to PyTorch-based technical stack and pays more attention to developers who have chosen PyTorch-related software packages. Finally, he/she chooses PyTorch (Torch) and PyTorch-Geomertic (TG, basically provides graph and tree analysis functions which are suitable for cell lineage analysis), and will further choose Torch-Cluster (TC) for cell lineage cluster analysis.

	\begin{table*}[htb]
		\centering
		\caption{The abbreviations and full names of software packages.}
		\label{tab:Abbreviation}
		\begin{tabular}{cc|cc}
			\toprule
			\multicolumn{2}{c|}{TensorFlow} & \multicolumn{2}{c}{PyTorch}\\
			\hline
			Abbreviation & Full Name &  Abbreviation & Full Name \\
			\midrule
			TF & TensorFlow & Torch & PyTorch \\
			DCTF & DeepCell-tf& TG & Torch-Geometric\\
			DCT & DeepCell-Tracking & TC & Torch-Cluster\\
			DCRM & DeepCell-RetinaMask & GC & GraphChem \\
			SP & Spektral & TP3d & Torch-Points3d \\
			TFT & TensorFlow-Text & TT & Torch-Text\\
			S2S & Seq2Seq & TS2S & PyTorch-Seq2Seq\\
			TFASR & TensorFlowASR & TSA & PyTorch-Sentiment-Analysis\\
			ONMT & OpenNMT & TGY & PyTorch-Geometric-YooChoose\\

			\bottomrule
		\end{tabular}
	\end{table*}
	
    In this article, we focus on modeling the dynamic interests of developers with both social influence and dependency constraints, and propose the Session-based Social and Dependency-aware software Recommendation (SSDRec) model. Our main contributions are summarized as follows:
    
    \begin{itemize}
        \item We propose to model the dynamic interests of developers with both social influence from friends and dependency constraints among software packages.

        \item We develop a unified framework to integrate a recurrent neural network and two graph attention networks. The recurrent neural network models the short-term dynamic interests of developers in each session and the two graph attention networks capture social influence from friends and dependency constraints from dependent software packages, respectively.

        \item We evaluate the proposed model on three real-world datasets from different communities in GitHub, including PHP, Ruby and JavaScript. The results demonstrate the performance and effectiveness of our proposed model compared with the baselines.
    \end{itemize}

    The rest of this article is organized as follows. In Section \ref{RELATED WORK}, we briefly review the important work related to this paper. Section \ref{PRELIMINARIES} provides some preliminaries of the article. Our proposed model SSDRec is described in Section \ref{METHODOLOGY} and the experiment results are shown in Section \ref{EXPERIMENTS}. Finally, we draw conclusions in Section \ref{CONCLUSIONS}.

    \section{Related work} \label{RELATED WORK}
    
    In this section, we will briefly review the related works in dynamic recommendation, social recommendation and software recommendation.
	
	\subsection{Dynamic recommendation}
	
    As users' interests dynamically evolve, conventional recommendation models which capture users' long-term static interests are not applicable anymore and various dynamic recommendation models have been proposed. For example, an earlier work utilized Markov Chains in successive point-of-interest recommendation \cite{cheng2013you}. With the success of RNN in several domains such as stock market prediction \cite{li2019multi} and sentiment analysis \cite{wang2018sentiment}, various approaches \cite{manotumruksa2017deep, dong2018recurrent} have been proposed to exploit Recurrent Neural Networks (RNN) to model users’ dynamic interests from their recent behaviors. Manotumruksa et al. \cite{manotumruksa2017deep} integrated the sequential time information of users' behaviors into the matrix factorization (MF) with an RNN. Dong et al. \cite{dong2018recurrent} further performed joint optimization of RNN and MF with shared parameters in a multi-task learning framework.
    
    Furthermore, instead of modeling behavior sequence as a whole as the above models, session-based recommendation models segment users' behavior sequence into several sessions to model users' dynamic interests in a more fine-grained granularity. Hidasi et al. \cite{hidasi2015session} first proposed an RNN-based approach for session-based recommendations to capture users' short-term dynamic interests within a session. Then based on the assumption that a session often serves different purposes, Wang et al. \cite{wang2019modeling} proposed a mixture-channel model with attention mechanism to detect the purposes of each item. In addition to modeling interactions within a session, information across sessions has been also introduced. Ruocco et al. \cite{ruocco2017inter} exploited two separate RNNs to process the current session and the past sessions separately.
	
	\subsection{Social recommendation}
	
	Social recommendation utilizes social network information to enhance the performance of recommendation models. Conventionally, this information is utilized by some hand-crafted features. Ma et al. \cite{2011Recommender} regularized the matrix factorization framework with social network information. Zhao et al. \cite{2014Leveraging} leveraged friends' interaction as another positive feedback for Bayesian Personalized Ranking (BPR). Ding et al. \cite{ding2018scfm} combined social information and crow computing into factorization machines for recommendation.
	
	Recently, deep neural network is adopted to process the social network information in recommendation models instead of hand-crafted features.  Deng et al. \cite{deng2016deep} utilized Autoencoder to initialize vectors in MF and updated them with both social trust ensemble and community effect. In order to extract structure information of social network more efficiently, graph neural network (GNN) is applied to recommendation systems \cite{wang2019neural}. Specifically, graph attention network \cite{veli2018graph} is usually utilized in recommendation systems to aggregate information from neighbors in a social network adaptively by learning different weights for different neighbors, which is also verified in rumor detection in social media \cite{bian2020rumor}. Fan et al. \cite{fan2019graph} coherently modeled two graphs and heterogeneous strengths to jointly capture interactions and opinions in the user--item graph. Song et al. \cite{song2019session} further considered the evolution of social influence and proposed DGRec to jointly utilize RNN and GAT to capture distinguishing social influence from different friends in different time period. Inspired by the idea behind DGRec and other social recommendation models of utilizing social influence to improve recommendation performance, we further extend this idea to a more generalized form. We consider the interdependent relations among not only users (i.e., social network) but also items (i.e., software dependency network), and propose a unified framework to extract information from both of these two interdependent relations in an end-to-end manner. Specifically, we apply our model to the field of software engineering where the items (i.e., software packages) are naturally interdependent and the interdependent relations are a key factor to affects developers' choices on software packages.
	
	\subsection{Software recommendation}
	
	Software recommendation emerges as a hot research field with the rapid development of social collaborative coding platforms such as GitHub and explosive growth of third-party software packages. Conventional software recommendation usually employed collaborative filtering-based recommendation models\cite{rendle2012bpr, he2017neural}. For example, Ichii et al. \cite{ichii2009software} used collaborative filtering to recommend similar software packages to developers. Thung et al. \cite{thung2013automated} further combined association rule and collaborative filtering to capture deeper relationships between software packages. He et al. \cite{he2020diversified} employed an adaptive weighting mechanism and neighborhood information to neutralize popularity bias in MF, which significantly increased both the diversity and accuracy of the recommendation results. In addition to directly reusing third-party software packages, Web service API is also reusable and recommendation models have been developed \cite{almarimi2019web,zhang2019location}.
	
	As social collaborative coding becomes popular and development stages change rapidly, some software recommendation models have begun to employ social influence or developers' dynamic behaviors. Guy et al. \cite{guy2009personalized} aggregated developers' familiarity network and similarity network to recommend social software. Jiang et al. \cite{jiang2020adapting} adopted time decay factor and operation behaviors weight to model developers' dynamic interests in a popular programming platform, Scratch. However, social influence or developers' dynamic behaviors are usually modeled separately and the dependency relations among software packages are usually ignored in existing works.
	
    \section{Preliminaries} \label{PRELIMINARIES}

    In this section, we will first introduce some necessary definitions and formulate the problem. The main notations are summarized in Table \ref{tab:notation}. Generally, sets, vectors and matrices are denoted as upper-case letters, bold lower-case letters and bold upper-case letters, respectively.
    
	\begin{table}
		\centering
		\caption{Summary of main notations}
		\label{tab:notation}
		\begin{tabular}{c|l}
			\toprule
			Symbol           & Description\\
			\midrule
			$U$              & The set of developers\\
			$I$              & The set of software packages\\
			$G_D$, $G_S$     & The dependency network and the social network \\
			$S_{T}^{u}$      & The session of developer $u$ during time period $T$\\
			$N_{D}(i)$       & $i$'s one-hop neighbors in dependency network\\
			$N_{S}(u)$       & $u$'s one-hop neighbors in social network\\
			$\alpha_{ij}^{(l)}$       & The attention weight between software package $i$ and $j$ in the $l$-th layer\\
			$\beta_{gf}^{(l)}$       & The attention weight between developer $g$ and $f$ in the $l$-th layer\\
			$\mathbf{h}_{t}$       & The hidden representation of LSTM at step $t$\\
			$\mathbf{e}_{i}^{(l)}, \mathbf{s}_{u}^{(l)}$ & the embeddings of software package $i$ and developer $u$ in $l$-th layer \\
			$\mathbf{e}_{i}, \mathbf{s}_{u,T}$   & the final embeddings of software package $i$ and developer $u$ \\
			$\mathbf{W}_{D}^{(l)}, \mathbf{W}_{S}^{(l)}$ & aggregation weight matrices for dependency and social network in $l$-th layer\\
			$\mathbf{W}_{TD}, \mathbf{W}_{TS}$ & transformation matrices for dependency and social network\\
			$p(i|S_{T}^{u})$ & The probability that developer $u$ chooses package $i$ in condition of session $S_{T}^{u}$\\
			\bottomrule
		\end{tabular}
	\end{table}

    \begin{definition}[Session]
    An ordered set of software packages with which the developer has interacted within a specific time period. Let $U$ and $I$ denote the sets of developers and software packages, respectively. The session of developer $u$ during time period $T$ is the ordered set of software packages developer $u$ watched within time period $T$, i.e., $S_{T}^{u} = <i_{T,1}^{u},i_{T,2}^{u}, \cdots ,i_{T,n}^{u}>$ ($u \in U$, $i_{T,k}^{u} \in I$ ).
    \end{definition}

    In the software development community, developers need to keep investigating new software packages as new development requirements arise or new techniques emerge. Especially, developers pay attention to different technical fields during different time periods. Thus, developers' historical interactions with software packages need to be segmented into different groups by time periods to represent their different interests in a fine-grained granularity and to capture the evolution of the interests.

    \begin{definition}[Dependency constraint]
    The dependency network $G_D=(I, E_D)$ formed by the dependency relations among software packages constrains developer $u$'s future choices to the near neighbors of his/her previous chosen software packages $\bigcup_{k=1}^{n} N_D(i_{T,k}^{u})$.
    \end{definition}
    
    As the complexity of modern software grows, it becomes a popular paradigm in modern software engineering to reuse third-party software packages. For example, to enable OAuth login using accounts from third-party social network platforms such as Facebook and Twitter, it is better to reuse well-tested third-party OAuth packages such as Laravel Socialite \footnote{https://github.com/laravel/socialite}, which will not only improve the efficiency of development but also guarantee the stability and security. Thus complicated dependency relations appear and form the dependency network. These dependency relations often constrain the technical choices of developers because developers often focus on specific technical fields which are in fact some sub-networks of the whole dependency network.

    \begin{definition}[Social influence]
    The choices of developer $u$'s friend developers $N_S(u)$ in the developer social network $G_S=(U, E_S)$ influence $u$'s future choices.
    \end{definition}
    
    Similar to the dependency relations among software packages, developers in social collaborative coding platforms like GitHub usually follow other developers and build the social network $G_S$. Then the friend developers' activities such as watching or forking a new software project appear in the developer's timeline, which will influence the future technical choices of developers.
    
    \begin{problem}[Session-based software recommendation]
    Given a new session $S_{T}^{u} = <i_{T,1}^{u}, \cdots ,i_{T,n}^{u}>$ of developer $u$, the goal of session-based software recommendation is to recommend a set of software packages from $I$ that the developer is most likely to watch in the next step $n+1$ within session $T$, i.e., $i_{T,n+1}^{u}$.
    \end{problem}
    
    In this article, we focus on the session-based software recommendation by simultaneously considering both dependency constraint among software packages and social influence among developers.
	
	\section{Methodology} \label{METHODOLOGY}
	
\begin{figure}[H]
		\centering
		\includegraphics[width=\linewidth]{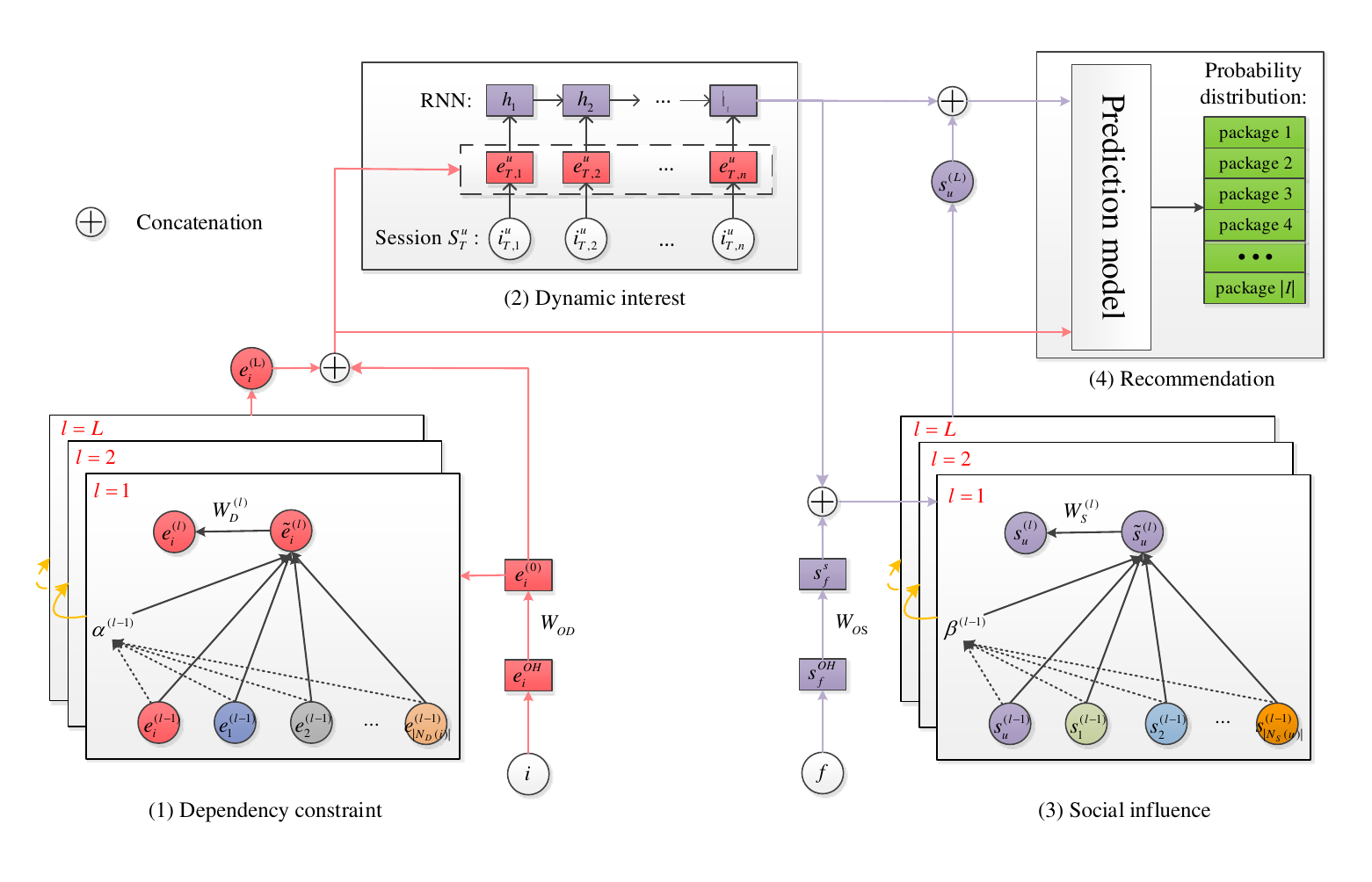}
		\caption{An overview of SSDRec.}
		\label{SSDRec}
	\end{figure}

    In this section, we present the proposed \textbf{S}ession-based \textbf{S}ocial and \textbf{D}ependency-aware software \textbf{Rec}ommendation model \textbf{SSDRec} in detail \footnote{The code and data are available at https://github.com/SSOTTY/SSDRec}. The overview of the model is shown in Figure \ref{SSDRec} and it is composed of four components: (1) Dependency constraint: we use a graph attention network to capture the dependency relations among software packages and obtain the embedding of each software package. (2) Dynamic interest modeling: we use a recurrent neural network to model the sequence of software packages in a session and obtain the final embedding of developer's dynamic interests in this session. (3) Social influence: we use another graph attention network to capture the social influence from neighbor developers and obtain the final embedding of each developer. (4) Recommendation: we use softmax to estimate the probability a developer will choose a given software package.
    
	\subsection{Dependency constraint}
	
    Dependency constraint can be regarded as the domain knowledge specific for software recommendation. We employ this domain knowledge into the embeddings of software packages by a graph attention network.
    
    First, the initial embedding of each software package is obtained by a transformation of its one-hot encoding,
    \begin{equation}
        \label{package-embedding-init}
		\mathbf{e}_{i}^{(0)} = \mathbf{W}_{OD} \cdot \mathbf{e}_{i}^{OH}
	\end{equation}
	where $\mathbf{e}_{i}^{OH} \in \mathbb{R}^{|I|}$ and $\mathbf{e}_{i}^{(0)} \in \mathbb{R}^{E}$ are the software package $i$'s one-hot encoding and initial embedding, respectively. $\mathbf{W}_{OD} \in \mathbb{R}^{E \times |I|}$ is the transformation matrix.
	
    Then the attention mechanism is utilized to distinguish the different dependency constraints from different neighbor software packages and the attention weight between software package $i$ and $j$ ($j \in N_{D}(i) \cup \{i\}$) in the $l$-th layer of the graph attention network is calculated as
	\begin{equation}
	    \label{dependency-attention-weight}
		\alpha_{ij}^{(l)}=\frac{exp((\mathbf{e}_{i}^{(l)})^{T} \cdot \mathbf{e}_{j}^{(l)})}{\sum_{k \in N_{D}(i)\cup \{i\}} exp((\mathbf{e}_{i}^{(l)})^{T} \cdot \mathbf{e}_{k}^{(l)})}
	\end{equation}
	where $\alpha_{ij}^{(l)} \in \mathbb{R}$ is the attention weight between software package $i$ and $j$, and $N_{D}(i)$ is the set of one-hop neighbors of software package $i$ in the dependency network. Note that we also use self-connection edge to preserve the embedding of software package $i$ itself.
	
	Finally, the embedding of software package $i$ is updated by a weighed aggregation of the embeddings of its neighbor packages from the previous layer,
	\begin{equation}
	    \label{dependency-embedding-aggregation}
		\mathbf{e}_{i}^{(l+1)}= \sigma (\mathbf{W}_{D}^{(l+1)}\sum_{j \in N_{D}(i)\cup \{i\}}\alpha_{ij}^{(l)}\mathbf{e}_{j}^{(l)})
	\end{equation}
	where $\mathbf{W}_{D}^{(l+1)} \in \mathbb{R}^{E \times E}$ is the aggregation weight matrix of the $(l+1)$-th layer and $\sigma$ is the nonlinear activation function.
	
	After $L$ layers' aggregation of dependency constraints, the embedding $\mathbf{e}_{i}^{(L)}$ is obtained. We then concatenate it with its original embedding $\mathbf{e}_{i}^{(0)}$ as the final embedding of the software package,
	\begin{equation}
	    \label{itemfinalrepresentation}
		\mathbf{e}_{i} = \mathbf{W}_{TD} \cdot (\mathbf{e}_{i}^{(0)} \parallel \mathbf{e}_{i}^{(L)})
	\end{equation}
	where $W_{TD} \in \mathbb{R}^{E \times 2E}$ is a linear transformation matrix.
	
	The overall procedure is summarized in Algorithm \ref{package-embedding}.
	
	\begin{algorithm}[H]
		\caption{Software package embedding with dependency constraints}
		\label{package-embedding}
		\LinesNumbered
		\KwIn{Software package set $I$; dependency network $G_D$; number of layers $L$}
		\KwOut{Embeddings of software packages $\mathbf{e}_{i}$ ( $i \in I$)}
		
		Initialize the embeddings of software packages according to Eq. (\ref{package-embedding-init});

		\For{l=0 \emph{\KwTo} L-1}{
			\ForEach{i in I}{
				\ForEach{j in $N_{D}(i)\cup {\{i\}}$}{
					Calculate the attention weight $\alpha_{ij}^{(l)}$ according to Eq.(\ref{dependency-attention-weight});
				}
				Obtain the embedding of $(l+1)$-th layer $\mathbf{e}_{i}^{(l+1)}$ according to Eq.(\ref{dependency-embedding-aggregation});
			}
		}
		
		\ForEach{i in I}{
		    Obtain the final embedding $\mathbf{e}_{i}$ according to Eq.(\ref{itemfinalrepresentation});
		}
	\end{algorithm}
	
	\subsection{Dynamic interest}
	
	Developers' interests gradually evolve as development requirements change and new technology emerges. We capture developers' dynamic interests in each session by modeling the sequence of software packages within each session using a recurrent neural network (RNN).
	
	Given the session $S_{T}^{u}=<i_{T,1}^{u},i_{T,2}^{u},i_{T,3}^{u},...,i_{T,n}^{u}>$ of developer $u$ in time period $T$, the embeddings of the software packages in this session $<e_{T,1}^{u},e_{T,2}^{u},e_{T,3}^{u},...,e_{T,n}^{u}>$ are first obtained from the previous step and then fed to an RNN. The RNN recurrently learns a hidden representation from the sequence by taking account both the current input package and previous input packages,
	\begin{equation}
		\mathbf{h}_{t} = RNN(\mathbf{e}_{T,t}^{u}, \mathbf{h}_{t-1})
	\end{equation}
	where $RNN$ is a kind of recurrent neural network model and $\mathbf{h}_{t} \in \mathbb{R}^{E} $ is the hidden representation with dimension $E$ at step $t$.
	
	Generally, there are various RNN models and we choose the long short-term memory (LSTM) model. Details of LSTM are shown in Equation (\ref{LSTM}),
    \begin{equation}
        \label{LSTM}
        \begin{aligned}
		& \mathbf{f}_{t} = \sigma(\mathbf{W}^{f}\mathbf{e}_{T,t}^{u}+\mathbf{U}^{f}\mathbf{h}_{t-1}+\mathbf{b}_{f})\\
		& \mathbf{i}_{t} = \sigma(\mathbf{W}^{i}\mathbf{e}_{T,t}^{u}+\mathbf{U}^{i}\mathbf{h}_{t-1}+\mathbf{b}_{i})\\
	    & \mathbf{o}_{t} = \sigma(\mathbf{W}^{o}\mathbf{e}_{T,t}^{u}+\mathbf{U}^{o}\mathbf{h}_{t-1}+\mathbf{b}_{o})\\
		& \mathbf{\hat{c}}_{t} = tanh(\mathbf{W}^{c}\mathbf{e}_{T,t}^{u}+\mathbf{U}^{c}\mathbf{h}_{t-1}+\mathbf{b}_{c})\\
		& \mathbf{c}_{t} = \mathbf{f}_{t}\odot\mathbf{c}_{t-1}+\mathbf{\hat{c}}_{t}\odot\mathbf{i}_{t}\\
		& \mathbf{h}_{t} = \mathbf{o}_{t}\odot tanh(\mathbf{c}_{t})
		\end{aligned}
	\end{equation}
	where $\mathbf{W} \in \mathbb{R}^{E \times E}$, $\mathbf{U} \in \mathbb{R}^{E \times E}$ and $\mathbf{b} \in \mathbb{R}^{E}$ are all model parameters, and $\odot$ denotes element-wise product.
	
	The overall procedure is summarized in Algorithm \ref{dynamic-interest-embedding}.
	
	\begin{algorithm}[H]
		\newpage
		\caption{Dynamic interest embedding}
		\label{dynamic-interest-embedding}
		\LinesNumbered
		\KwIn{Session $S_{T}^{u}=<i_{T,1}^{u},i_{T,2}^{u},i_{T,3}^{u},...,i_{T,n}^{u}>$ of developer $u$ in time period $T$; embeddings of software packages $\mathbf{e}_{i}$ ( $i \in I$)}
		\KwOut{Dynamic interest embedding $\mathbf{s}_{u,T}^{d}$ of developer $u$ in session $T$}
		
		\For{t=1 \emph{\KwTo} $|S_{T}^{u}|$}{
		    Calculate the hidden representation of LSTM  $\mathbf{h}_t$ according to Eq. (\ref{LSTM});
		}
		
		Assign the final hidden representation of LSTM $\mathbf{h}_{|S_{T}^{u}|}$ as the dynamic interest embedding of developer $u$ in session $T$, $\mathbf{s}_{u,T}^{d} \leftarrow \mathbf{h}_{|S_{T}^{u}|}$;
		
	\end{algorithm}

	\subsection{Social influence}
	
	In social collaborative coding platforms like GitHub, a developer can follow other developers to have their recent activities appeared in his/her timeline. Thus, neighbor developers' choices can influence the developer's own choice. We capture this social influence by applying a graph attention network on the social network of developers.

	First, for a given developer $u$ at time period $T$, the initial embedding of each of his/her neighbor $f \in N_S(u)$ is obtained by a combination of his/her one-hot encoding and dynamic interest of previous time period $T-1$,
	\begin{equation}
        \label{developer-embedding-init}
		\mathbf{s}_{f}^{(0)} = \sigma (\mathbf{W}_{f} (\mathbf{s}_{f,T-1}^{d} \parallel \mathbf{W}_{OS} \cdot \mathbf{s}_{f}^{OH}))
	\end{equation}
	where $\mathbf{s}_{f}^{OH} \in \mathbb{R}^{|U|}$, $\mathbf{s}_{f,T-1}^{d} \in \mathbb{R}^{E}$ and $\mathbf{s}_{f}^{(0)} \in \mathbb{R}^{E}$ are the developer $f$'s one-hot encoding, dynamic interest embedding at time period $T-1$ and initial embedding, respectively. $\mathbf{W}_{f} \in \mathbb{R}^{E \times 2E}$ and $\mathbf{W}_{OS} \in \mathbb{R}^{E \times |U|}$ are the transformation matrices. The reason why neighbors' dynamic interest of previous time period is utilized comes from the practical situation that a session has a certain length (e.g., one week) while within a certain session $T$, developers’ activities usually happen at a time point in a finer granularity (e.g., day). For example, in session $T$, the last activity of developer $u$ happens on Wednesday and to predict his/her preference in the following days in this session, it not reasonable to involve his/her neighbor $f$’s activity which will happen in the future, e.g., Friday, in the modeling of social influence. Thus, we model the social influence from $T=2$ with neighbors’ dynamic interest embeddings from the last session $T-1$ ($T-1 \geq 1$).

    Specially, for the target developer $u$, we initialize his/her embedding using the dynamic interest embedding at time period $T$,
    \begin{equation}
        \label{developer-u-embedding-init}
		\mathbf{s}_{u}^{(0)} = \mathbf{s}_{u,T}^{d}
	\end{equation}
	
	Then similar to dependency constraints, the attention mechanism is also utilized to distinguish the different social influence from different neighbor developers. The attention weight between developer $g$ and $f$ ($f \in N_{S}(g)  \cup \{g\}$) in the $l$-th layer of the graph attention network is calculated as
	\begin{equation}
	    \label{social-attention-weight}
		\beta_{gf}^{(l)}=\frac{exp((\mathbf{s}_{g}^{(l)})^{T} \cdot \mathbf{s}_{f}^{(l)})}{\sum_{k \in N_{S}(g) \cup \{g\}} exp((\mathbf{s}_{g}^{(l)})^{T} \cdot \mathbf{s}_{k}^{(l)})}
	\end{equation}
	where $\beta_{gf}^{(l)} \in \mathbb{R}$ is the attention weight between developer $g$ and $f$, and $N_{S}(g)$ is the set of one-hop neighbors of developer $g$ in the social network.
	
	Finally, the embedding of developer $g$ at time period $T$ is updated by a weighed aggregation of the embeddings of its neighbor developers from the previous layer,
	\begin{equation}
	    \label{social-embedding-aggregation}
		\mathbf{s}_{g}^{(l+1)}= \sigma (\mathbf{W}_{S}^{(l+1)} \sum_{f \in N_{S}(g) \cup \{g\}}\beta_{gf}^{(l)}\mathbf{s}_{f}^{(l)})
	\end{equation}
	where $\mathbf{W}_{S}^{(l+1)} \in \mathbb{R}^{E \times E}$ is the aggregation weight matrix of the $(l+1)$-th layer and $\sigma$ is the nonlinear activation function.
	
	After $L$ layers' aggregation of social influence, the embedding  $\mathbf{s}_{u}^{(L)}$ is obtained. We then concatenate it with its initial embedding $\mathbf{s}_{u}^{(0)}$ as the final embedding of developer $u$,
	\begin{equation}
	    \label{userfinalrepresentation}
		\mathbf{s}_{u,T} = \mathbf{W}_{TS} \cdot (\mathbf{s}_{u}^{(0)} \parallel \mathbf{s}_{u}^{(L)})
	\end{equation}
	where $W_{TS} \in \mathbb{R}^{E \times 2E}$ is a linear transformation matrix.
	
	The overall procedure is summarized in Algorithm \ref{developer-embedding}.
	
	\begin{algorithm}[H]
		\caption{Developer embedding with social influence}
		\label{developer-embedding}
		\LinesNumbered
		\KwIn{Target developer $u$; time period $T$; session set $S=\{S_{T}^{u}\}$; social network $G_S$; neighborhood sampling functions $N_{S}^{(l)}$; number of layers $L$}
		\KwOut{Developer $u$'s embedding $\mathbf{s}_{u,T}$}
		
		Initialize neighbor set $B^{L} \leftarrow \{u\}$;
		
		\For{k=L \emph{\KwTo} 1}{
		    Initialize neighbor set $B^{k-1} \leftarrow B^{k}$;
		    
		    \ForEach{g in $B^{k}$}{
		        Sample neighbors of $g$ using neighborhood sampling function $N_{S}^{(k)}(g)$ and add to neighbor set $B^{k-1} \leftarrow B^{k-1} \cup N_{S}^{(k)}(g)$;
		    }
		}
		
		\ForEach{g in $B^{0}$}{
            Obtain the initial embedding $\mathbf{s}_{g}^{(0)}$ of $g$ according to Eq. (\ref{developer-embedding-init}) and Eq. (\ref{developer-u-embedding-init});
		}
		
		\For{k=1 \emph{\KwTo} L}{
		    \ForEach{g in $B^{k}$}{
		        Calculate attention weight between $g$ and its neighbors according to Eq. (\ref{social-attention-weight});
		        
		        Calculate the embedding $\mathbf{s}_{g}^{(k)}$ of $g$ in $k$-th layer according to Eq. (\ref{social-embedding-aggregation});
		    }
		}
		
		Calculate the final embedding $\mathbf{s}_{u,T}$ of developer $u$ according Eq. (\ref{userfinalrepresentation});
	\end{algorithm}

	\subsection{Recommendation}
	
	After obtaining the embeddings of the software packages and the developer $u$, we employ the softmax to estimate the probability that developer $u$ will choose each software package $i$ and recommend the top-$K$ software packages to developer $u$.
	\begin{equation}
	    \label{recommendation}
		p(i|S_{T}^{u})=\frac{exp(\mathbf{e}_{i}^{T} \cdot \mathbf{s}_{u,T})}{\sum_{z=1}^{|I|}exp(\mathbf{e}_{z}^{T} \cdot \mathbf{s}_{u,T})}
	\end{equation}
	
	\subsection{Optimization}
	
	To learn the model parameters, we maximize the log-likelihood of the observed packages in all sessions,
	\begin{equation}
		\sum_{u\in U}\sum_{t=2}^{T}\sum_{n=1}^{L_{u,t}-1}log(p(i^{u}_{t,n+1}|i^{u}_{t,1},...,i^{u}_{t,n}))
	\end{equation}
	where $L_{u,t}$ denotes the lenghth of session $t$ of developer $u$. As our model is an end-to-end framework, all components (i.e., two GATs and an RNN) are combined into a single neural network model for optimization. In particular, we optimize the model parameters by utilizing the widely-adopted data augmentation and mini-batch technique. Specifically, we first make the first $m$ items of a session as a new session and then combine every $M$ sessions into a batch for a round of parameter updating.
	
	The overall training procedure is summarized in Algorithm \ref{model-training}.
    
    \begin{algorithm}[H]
		\caption{Model training}
		\label{model-training}
		\LinesNumbered
		\KwIn{Developer set $U$; software package set $I$; dependency network $G_D$; social network $G_S$; training set $S=\{S_{T}^{u}\}$; mini-batch size $M$}
		\KwOut{Model parameters $\Theta$}
		
		Obtain the embedding of each software package according to Algorithm \ref{package-embedding};

		\ForEach{$M$ sessions in $S$ as a batch $BT$}{
		
		    Set augmented session datasets $SA=\{\}$;
		
            \ForEach{$S_{T}^{u}$ in $BT$}{
		        \For{m=1 \emph{\KwTo} $|S_{T}^{u}|-1$}{
		            Obtain the first $m$ items of $S_{T}^{u}$ and make a new session $S'=<i_{T,1}^{u}, \cdots ,i_{T,m}>$;
		        
		           Add $S'$ to $SA$;
		        }
	        }
		
		    Set probability $P=0$
		    
		    \ForEach{$S'$ in $SA$}{
		        Obtain the embedding of developer $u$ from session $S'$ according to Algorithm \ref{developer-embedding};
		        
		        Calculate the possibility $p(i_{T,m+1}^{u}|S')$ that developer $u$ chooses $i_{T,m+1}^{u}$ according to Eq.(\ref{recommendation});
		        
		        Add $p(i_{T,m+1}^{u}|S')$ to $P$;
		    }
		    
		    Update model parameters $\Theta$ by maximizing the log-likelihood of $P$;
		}
	\end{algorithm}

	\section{Experiments} \label{EXPERIMENTS}
	
	In this section, we conduct experiments to answer the following research questions:
	\begin{itemize}
		\item{RQ1} Does SSDRec outperform the baseline methods on all experimental settings?
		\item{RQ2} Do the components of SSDRec enhance the effectiveness by (a) modeling social influence and dependency constraints, (b) capturing friends' dynamic and static interests. 
		\item{RQ3} How do some hyper-parameters affect the performance of SSDRec?
	\end{itemize}
	
	In the remainder of this section, we will first describe the experimental settings (Section \ref{experimental-settings}) and then answer the above research questions(Section \ref{Comparison}, Section \ref{Ablation}, Section \ref{parameters}). Finally, several illustrative examples are given to verify our assumptions.(Section \ref{Visualization}).
	
	\subsection{Experimental Settings} \label{experimental-settings}
	
	\subsubsection{Datasets}
	
    To evaluate the proposed approach, we integrate the datasets of GHTorrent \cite{Gousi13} and Libraries.io \footnote{https://libraries.io/data}. GHTorrent monitors the public event timeline of Github and provides abundant social relations between developers and development interactions between developers and software repositories from the popular social collaborative coding platform GitHub. And Libraries.io provides the explicit dependency relations between software packages. In our experiments, we extract three datasets from the integration of GHTorrent and Libraries.io according to the programming language, i.e., PHP, Ruby and JavaScript. For each extracted dataset, the social relations between developers, the watch interactions between developers and software repositories, and the dependency relations between software packages are extracted. To ensure enough social awareness for each developer, we only extract those developers who follow more than $m$ developers and are followed by more than $n$ developers. Similarly, we keep the software packages that are watched by at least $k$ developers. Due to the different popularity of those three programming languages, we set $m=20, 50, 100$, $n=50, 100, 200$ and $k=20, 20, 50$ for PHP, Ruby and JavaScript, respectively. Then the watched items of each developer are divided into week-long sessions and only those sessions whose length is within $[2, 30]$ are kept according to \cite{song2019session}.
    
    \begin{table}[H]
    	\centering
		\caption{The statistics of datasets.}
		\label{tab:dataset}
		\begin{tabular}{lccc}
			\toprule
			Dataset&PHP&Ruby&JavaScript\\
			\midrule
			$\#$ Developers & 7,095 & 5,935 & 8,144\\
			$\#$ Software packages & 1,402 & 1,487 & 2,282\\
			$\#$ Interactions & 118,256 & 116,832 & 436,409\\
			\hline
			Avg. friends per developer & 8.59 & 13.05 & 16.24\\
			Avg. dependencies per package & 4.56 & 5.09 & 10.09\\
			Avg. session length & 2.95 & 3.13 & 3.25\\
			\bottomrule
		\end{tabular}
	\end{table}

    Each dataset is then split into training, validation and test sets. Sessions of each developer are ordered by time in ascending order and the sessions within the last two years (about 104 weeks) are randomly split into validation and test sets. The rest sessions are used as training set. Note that when splitting datasets, we ensure all packages in validation/test sets appear in the training set. The detailed statistics of these three datasets are shown in Table \ref{tab:dataset}.

	\subsubsection{Baselines}
    
    To evaluate the performance of our proposed SSDRec, we compare it with the following baselines. As software recommendation can utilize conventional recommendation models with specific features, we choose the two classic recommendation modes, i.e., BPR \cite{rendle2012bpr} and NCF \cite{he2017neural} as baselines. Moreover, considering the dynamic nature we also choose two session-based recommendation models, i.e., RNN \cite{hidasi2015session} and DGRec \cite{song2019session}. As for recommendation models specifically designed for software recommendation, because they usually focuses on enhancing recommendation performance by incorporating additional meta information, such as tag and domain-specific knowledge \cite{chen2016similartech}, programming language \cite{jiang2017open}, description files \cite{sun2018personalized}, textual requirement descriptions \cite{sun2020req2lib} and multiple developer--software interactions \cite{zhang2018funkr} and there is hardly any software recommendation models utilizing the natural dependency relations among software packages except RepoLike \cite{yang2019repolike}. We also choose RepoLike as a baseline. A brief introduction of these baselines are depicted as follows:
	\begin{itemize}
		\item BPR \cite{rendle2012bpr}: a classical MF-based method optimized with a ranking objective.
		\item NCF \cite{he2017neural}: uses neural network instead of inner product to model relationship between users and items of MF.
		\item ReopLike \cite{yang2019repolike}:a personalized recommendation approach that uses project popularity, technical dependencies and social connections for open-source repositories.
		\item RNN \cite{hidasi2015session}: captures users' session-level dynamic interests with RNN.
		\item DGRec \cite{song2019session}: a state-of-the-art model for session-based social recommendation. It utilizes RNN and GAT to model dynamic interests and dynamic social influence.
	\end{itemize}
	
	\subsubsection{Parameter Settings}
	
	Our proposed model is implemented using TensorFlow \cite{abadi2016tensorflow}. Adam \cite{kingma2014adam} is chosen as optimizer and parameters are initialized as suggested in \cite{abadi2016tensorflow}. The batch size and dropout are set to 200 and 0.2, respectively. The embedding dimensions of both developers and software packages for all models are set to 100. For both social network and dependency network, we employ the graph attention networks with 2 layers, i.e., $L=2$ according to \cite{hamilton2017inductive} and for each node 10 one-hop neighbors and 5 two-hop neighbors are sampled as recommended in \cite{hamilton2017inductive}. All experiments are run on a machine with a GeForce RTX2080Ti GPU.
	
	\subsubsection{Evaluation Metrics}
	
	The performance of SSDRec and all baselines are evaluated with two well-known metrics: Hit Rate $HR@K$ and Normalized Discounted Cumulative Gain $NDCG@K$.
	
    $HR@K$ measures the proportion of users who get correct recommendation results and it can be formulated as
	\begin{equation}
		HR@K=\frac{\sum_{u \in U} I(R_{K}(u) \cap T(u))}{|U|}
	\end{equation}
	where $R_{K}(u)$ and $T(u)$ are the sets of top-K recommendation list and ground truth list for user $u$, respectively. $I(\cdot)$ is the indicator function and $I(R_{K}(u) \cap T(u)) = 0$ when $R_{K}(u) \cap T(u) = \emptyset$ otherwise it is 1. |U| is the total number of users.
	
    $NDCG@K$ considers the orders of users' preferred items in the ranked Top-N recommendation list for users usually pay attention to only the top few items recommended by a recommendation system. It can be formulated as:
	\begin{equation}
		NDCG@K=\frac{1}{|U|}\sum_{u \in U}\frac{1}{log_{2}(1+pos(T(u)))}
	\end{equation}
	where $pos(T(u))$ is the order of the top ranked item of $u$'s ground truth items in the recommendation list.
	
	\subsection{Overall performance (RQ1)} \label{Comparison}
	
	\begin{table*}[htb]
	\tiny
		\centering
		\caption{The overall performance. The Improvement rate is calculated between the best and the second best performed models.}
		\label{tab:comparision}
		\begin{tabular}{cccccccc}
			\toprule
			\multirow{2}{*}{Dataset} & \multirow{2}{*}{Model} & \multicolumn{3}{c}{$HR@K$($\%$)} & \multicolumn{3}{c}{$NDCG@K$($\%$)}\\
			&& 10 & 20 & 50 & 10 &  20 & 50\\
			\midrule
			\multirow{6}{*}{PHP} &
			BPR & 3.41$\pm$0.159 & 6.70$\pm$0.104  & 15.26$\pm$0.111 & 1.40$\pm$0.064 & 2.22$\pm$0.048 & 3.90$\pm$0.050 \\
			& NCF & 4.50$\pm$0.054 & 8.13$\pm$0.040  & 18.35$\pm$0.161 & 2.08$\pm$0.025 & 2.99$\pm$0.002 & 5.00$\pm$0.031 \\
			& RepoLike & 5.54$\pm$0.111 & 8.65$\pm$0.034  & 15.02$\pm$0.055 & 2.76$\pm$0.228 & 3.59$\pm$0.026 & 4.86$\pm$0.092 \\
			& RNN & 9.75$\pm$0.099 & 16.23$\pm$0.132  & 28.72$\pm$0.238 & 4.55$\pm$0.045 & 6.22$\pm$0.033 & 8.76$\pm$0.047 \\
			& DGRec & 11.07$\pm$0.040 & 17.29$\pm$0.020 & 29.64$\pm$0.069 & 5.37$\pm$0.024 & 7.04$\pm$0.025 & 9.55$\pm$0.015 \\
			& SSDRec & \textbf{11.95$\pm$0.081} & \textbf{18.93$\pm$0.135} & \textbf{31.13$\pm$0.077} & \textbf{5.88$\pm$0.042} & \textbf{7.74$\pm$0.020} & \textbf{10.22$\pm$0.005} \\
			\hline
			& Improvement rate (\%) & 7.97 & 9.44 & 5.02 & 9.49 & 9.96 & 6.98 \\
			\hline
			\multirow{6}{*}{Ruby} &
			BPR & 2.09$\pm$0.155 & 4.13$\pm$0.200  & 10.16$\pm$0.094 & 0.88$\pm$0.086 & 1.39$\pm$0.098 & 2.56$\pm$0.037 \\
			& NCF & 2.83$\pm$0.098 & 5.03$\pm$0.099  & 11.17$\pm$0.088 & 1.34$\pm$0.041 & 1.88$\pm$0.037 & 3.09$\pm$0.029  \\
			& RepoLike & 5.97$\pm$0.331 & 10.15$\pm$0.063  & 17.54$\pm$0.045 & 3.11$\pm$0.201 & 4.17$\pm$0.025 & 5.69$\pm$0.074 \\
			& RNN & 6.25$\pm$0.071 & 10.07$\pm$0.038  & 19.17$\pm$0.075 & 3.18$\pm$0.052 & 4.19$\pm$0.078 & 5.98$\pm$0.069\\
			& DGRec & 7.00$\pm$0.027 & 10.86$\pm$0.027 & 20.73$\pm$0.058 & 3.68$\pm$0.006 & 4.70$\pm$0.001 & 6.68$\pm$0.017 \\
			& SSDRec & \textbf{7.63$\pm$0.169} & \textbf{12.05$\pm$0.072} & \textbf{21.98$\pm$0.115} & \textbf{4.01$\pm$0.047} & \textbf{5.19$\pm$0.039} & \textbf{7.17$\pm$0.045} \\
			\hline
			& Improvement rate (\%) & 9.08 & 10.96 & 6.02 & 8.93 & 10.40 & 7.43 \\
			\hline
			\multirow{6}{*}{JavaScript} &
			BPR & 1.92$\pm$0.057 & 3.59$\pm$0.074  & 7.74$\pm$0.160 & 0.85$\pm$0.029 & 1.27$\pm$0.025 & 2.08$\pm$0.019 \\
			& NCF & 2.32$\pm$0.012 & 4.11$\pm$0.130  & 8.42$\pm$0.107 & 1.07$\pm$0.029 & 1.52$\pm$0.058 & 2.37$\pm$0.054 \\
			& RepoLike & 2.50$\pm$0.029 & 4.40$\pm$0.037  & 8.53$\pm$0.039 & 1.13$\pm$0.026 & 1.64$\pm$0.017 & 2.46$\pm$0.090 \\
			& RNN & 4.59$\pm$0.021 & 7.91$\pm$0.100  & 15.27$\pm$0.102 & 2.19$\pm$0.005 & 3.06$\pm$0.029 & 4.53$\pm$0.024 \\
			& DGRec & 5.26$\pm$0.046 & 8.87$\pm$0.061 & 17.58$\pm$0.065 & 2.53$\pm$0.005 & 3.47$\pm$0.024 & 5.22$\pm$0.031 \\
			& SSDRec & \textbf{5.70$\pm$0.085} & \textbf{9.73$\pm$0.149} & \textbf{18.47$\pm$0.282} & \textbf{2.65$\pm$0.034} & \textbf{3.73$\pm$0.049} & \textbf{5.48$\pm$0.076}\\
			\hline
			& Improvement rate (\%) & 8.32 & 9.71 & 5.05 & 5.02 & 7.52 & 4.92 \\
			\bottomrule
		\end{tabular}
	\end{table*}
	
	The overall performance of our proposed SSDRec and all the baselines on the three real world datasets is shown in Table \ref{tab:comparision} from which we can obtain the following observations: 1) SSDRec significantly outperforms all the baselines with at least $4.92\%$ improvement over the second best model. 2) Static models BPR and NCF have much worse performance indicating the interests of developers change frequently and it is of great importance to model the dynamics of developers' interests. 3) Models considering social influence, i.e., DGRec and SSDRec have a better performance which enhances the assumption that the social relations among developers have an important impact on the evolution of developers' interests. 4) The significant improvement of SSDRec over DGRec also indicates the important impact of dependency constraints which agrees with the practices in software development, that is, a developer usually chooses to watch new software packages based on his/her existing technical stacks. 5) SSDRec is best suited to the PHP dataset. The reason may be that PHP is a dedicated programming language for backend web development while Ruby and JavaScript have diversified application fields, and in the PHP community, developers have tight social and dependency relations and shorter session length.
	
	\subsection{Ablation Studies (RQ2)} \label{Ablation}
	
    As SSDRec is composed of several components, we conduct the ablation studies by comparing the performance of the variants of SSDRec to demonstrate the effectiveness of different components.
	
	\subsubsection{Effect of social network and dependency network (RQ2(a))}
    
    SSDRec utilizes two graph attention networks to capture the social influence among developers and dependency constraints among software packages. To illustrate their impact on the recommendation performance, we compare the performance of the two variants of SSDRec, i.e., SSDRec-social and SSDRec-dependency with SSDRec. SSDRec-social and SSDRec-dependency exclude the graph attention network for dependency network and social network, respectively. Actually, SSDRec-social only considers the social influence and is identical to DGRec. The detailed modifications of SSDRec variants are shown in Table \ref{tab:variantofnetwork}.
	
	\begin{table}[H]
		\centering
		\caption{Detailed modifications of SSDRec variants. $W_{T}$ is the transformation matrix to transform the dimension of developer to $E$.}
		\label{tab:variantofnetwork}
		\begin{tabular}{l c}
			\toprule
			Variant & Modification\\
			\midrule
			SSDRec-social & Eq.(\ref{itemfinalrepresentation}) $\rightarrow$ $\mathbf{e}_{i}=\mathbf{e}_{i}^{(0)}$ \\
			SSDRec-dependency & Eq.(\ref{userfinalrepresentation}) $\rightarrow$ $\mathbf{s}_{u,T}=W_{T} \cdot \mathbf{s}_{u}^{(0)}$ \\
			\bottomrule
		\end{tabular}
	\end{table}
	
	\begin{table}[H]
	    \tiny
		\centering
		\caption{Performance of SSDRec and its variants concerning about social influence or dependency constraints.}
		\label{tab:comparisionofnetwork}
		\centering
		\begin{tabular}{cccccccc}
			\toprule
			\multirow{2}{*}{Dataset} & \multirow{2}{*}{Model} & \multicolumn{3}{c}{$HR@K$($\%$)} & \multicolumn{3}{c}{$NDCG$@K($\%$)}\\
			&& 10 & 20 & 50 & 10 & 20 & 50\\
			\midrule
			\multirow{3}{*}{PHP} &
			SSDRec-social & 11.07$\pm$0.040 & 17.29$\pm$0.020 & 29.64$\pm$0.069 & 5.37$\pm$0.024 & 7.04$\pm$0.025 & 9.55$\pm$0.015 \\
			& SSDRec-dependency & 11.07$\pm$0.061 & 17.70$\pm$0.043 & 29.89$\pm$0.080 & 5.30$\pm$0.036 & 7.05$\pm$0.042 & 9.53$\pm$0.034 \\
			& SSDRec & \textbf{11.95$\pm$0.081} & \textbf{18.93$\pm$0.135} & \textbf{31.13$\pm$0.077} & \textbf{5.88$\pm$0.042} & \textbf{7.74$\pm$0.020} & \textbf{10.22$\pm$0.005} \\
			\hline
			\multirow{3}{*}{Ruby} &
			SSDRec-social & 7.00$\pm$0.027 & 10.86$\pm$0.027 & 20.73$\pm$0.058 & 3.68$\pm$0.006 & 4.70$\pm$0.001 & 6.68$\pm$0.017 \\
			& SSDRec-dependency & 7.25$\pm$0.111 & 10.69$\pm$0.099 & 19.76$\pm$0.473 & 3.74$\pm$0.024 & 4.65$\pm$0.025 & 6.44$\pm$0.109 \\
			& SSDRec & \textbf{7.63$\pm$0.169} & \textbf{12.05$\pm$0.072} & \textbf{21.98$\pm$0.115} & \textbf{4.01$\pm$0.047} & \textbf{5.19$\pm$0.039} & \textbf{7.17$\pm$0.045} \\
			\hline
			\multirow{3}{*}{JavaScript} &
			SSDRec-social & 5.26$\pm$0.046 & 8.87$\pm$0.061 & 17.58$\pm$0.065 & 2.53$\pm$0.005 & 3.47$\pm$0.024 & 5.22$\pm$0.031 \\
			& SSDRec-dependency & 4.68$\pm$0.072 & 8.04$\pm$0.037 & 15.73$\pm$0.071 & 2.28$\pm$0.064 & 3.14$\pm$0.046 & 4.69$\pm$0.058 \\
			& SSDRec & \textbf{5.70$\pm$0.085} & \textbf{9.73$\pm$0.149} & \textbf{18.47$\pm$0.282} & \textbf{2.65$\pm$0.034} & \textbf{3.73$\pm$0.049} & \textbf{5.48$\pm$0.076}\\
			\bottomrule
		\end{tabular}
	\end{table}
	
	The results are shown in Table \ref{tab:comparisionofnetwork}. SSDRec has better performance than both of its variants, which validates the effectiveness of the composition of both components in our model. Moreover, it can also be found that social influence and dependency constraints have different impacts in different development communities. In PHP community, SSDRec-dependency usually outperforms SSDRec-social while in Ruby and JavaScript communities it is just the opposite.

	\subsubsection{Effect of dynamic interests and static interests (RQ2(b))}
	
    In the software development community, a developer is generally specialized in a specific technical base and gradually expand it with newer technique. For example, a frontend developer is usually specialized in JavaScript and gradually learns different frontend JavaScript frameworks as they are evolving rapidly, from jQuery to Angular to Vue.js. Thus, developers' interests can be divided into long-term static interests and short-term dynamic interests. Table \ref{tab:comparisionofprefrence} shows the effectiveness of capturing both dynamic and static interests for software recommendation. Especially, in PHP community and JavaScript community dynamic interests play a more important role, which is the opposite in Ruby community. This is because PHP community and JavaScript community evolve more rapidly and new techniques and frameworks keep emerging all the time.

	\begin{table}[H]
	    \tiny
		\caption{Performance of SSDRec and its variants concerning about dynamic or static interests. }
		\label{tab:comparisionofprefrence}
		\centering
		\begin{tabular}{cccccccc}
			\toprule
			\multirow{2}{*}{Dataset} & \multirow{2}{*}{Model} & \multicolumn{3}{c}{HR@K($\%$)} & \multicolumn{3}{c}{NDCG@K($\%$)}\\
			&& 10 & 20 & 50 & 10 & 20 & 50\\
			\midrule
			\multirow{3}{*}{PHP} &
			SSDRec-dynamic & 11.37$\pm$0.011 & 18.04$\pm$0.098 & 30.83$\pm$0.202 & 5.68$\pm$0.014 & 7.46$\pm$0.047 & 10.05$\pm$0.056 \\
			& SSDRec-static & 11.03$\pm$0.036 & 17.62$\pm$0.109 & 30.00$\pm$0.055 & 5.44$\pm$0.028 & 7.23$\pm$0.031 & 9.73$\pm$0.013 \\
			& SSDRec & \textbf{11.95$\pm$0.081} & \textbf{18.93$\pm$0.135} & \textbf{31.13$\pm$0.077} & \textbf{5.88$\pm$0.042} & \textbf{7.74$\pm$0.020} & \textbf{10.22$\pm$0.005} \\
			\hline
			\multirow{3}{*}{Ruby} &
			SSDRec-dynamic & 7.30$\pm$0.527 & 11.34$\pm$0.565 & 21.29$\pm$0.416 & 3.77$\pm$0.265 & 4.81$\pm$0.270 & 6.82$\pm$0.227 \\
			& SSDRec-static & 7.39$\pm$0.296 & 11.36$\pm$0.085 & 21.01$\pm$0.099 & 3.84$\pm$0.067 & 4.88$\pm$0.315 & 6.81$\pm$0.086 \\
			& SSDRec & \textbf{7.63$\pm$0.169} & \textbf{12.05$\pm$0.072} & \textbf{21.98$\pm$0.115} & \textbf{4.01$\pm$0.047} & \textbf{5.19$\pm$0.039} & \textbf{7.17$\pm$0.045} \\
			\hline
			\multirow{3}{*}{JavaScript} &
			SSDRec-dynamic & 5.54$\pm$0.169 & 9.50$\pm$0.038 & 18.23$\pm$0.068 & 2.58$\pm$0.005 & 3.62$\pm$0.032 & 5.37$\pm$0.005 \\
			& SSDRec-static & 5.10$\pm$0.042 & 8.67$\pm$0.031 & 16.76$\pm$0.195 & 2.36$\pm$0.003 & 3.30$\pm$0.038 & 4.92$\pm$0.034 \\
			& SSDRec & \textbf{5.70$\pm$0.085} & \textbf{9.73$\pm$0.149} & \textbf{18.47$\pm$0.282} & \textbf{2.65$\pm$0.034} & \textbf{3.73$\pm$0.049} & \textbf{5.48$\pm$0.076}\\
			\bottomrule
		\end{tabular}
	\end{table}
	
    \subsection{Hyperparameter Analysis (RQ3)} \label{parameters}

    To ensure the flexibility of the proposed SSDRec, it employs several hyperparameters. In this section, we conduct experiments to show how the hyperparameters affect the performance of our model.
	
	\subsubsection{Neighborhood sample size}

\begin{figure}[H]
		\centering
		\subfigure[PHP-social]{
			\label{influenceofnneighbors:c}
			\includegraphics[width=0.46\linewidth]{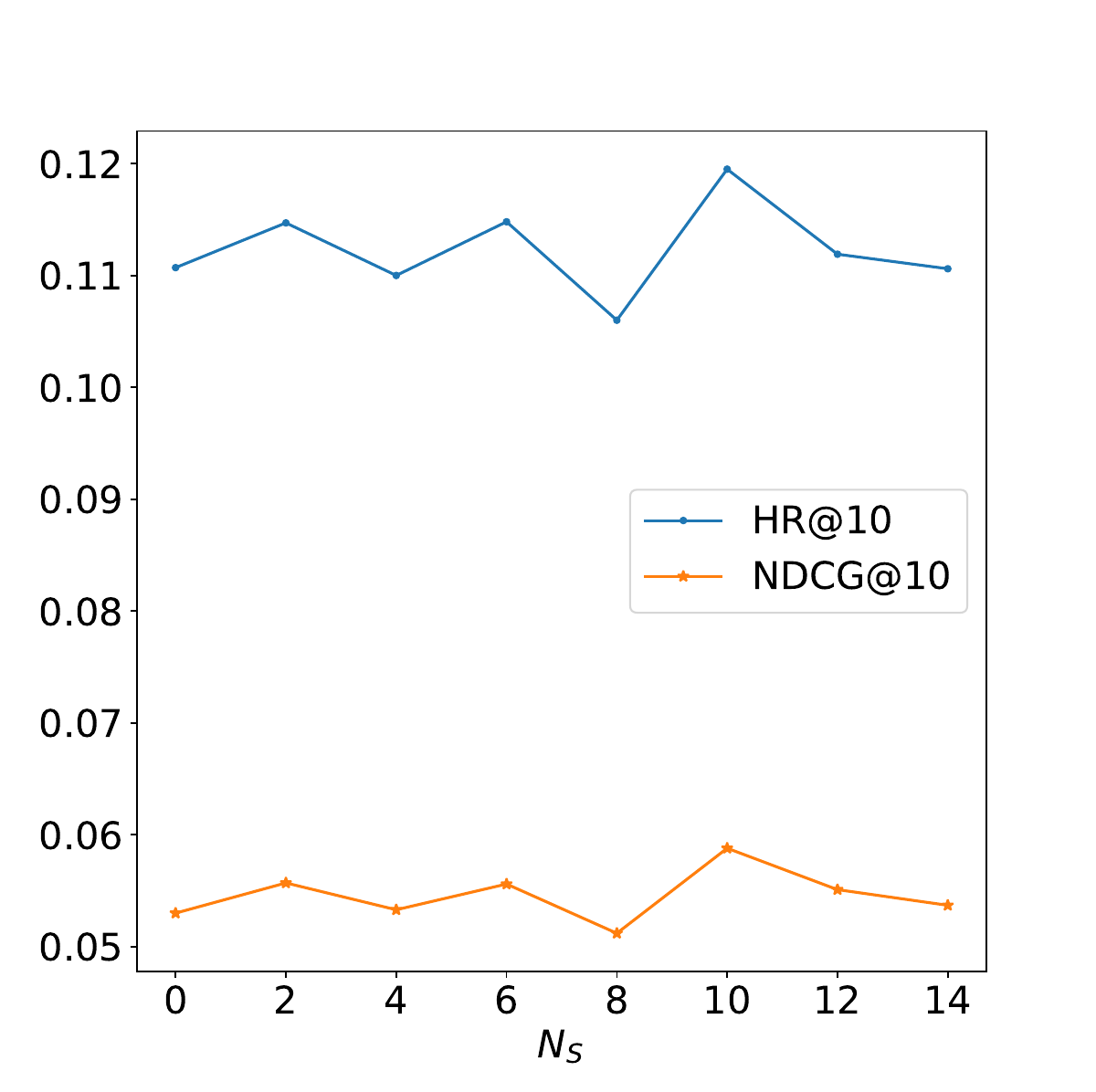}}
		\subfigure[Ruby-social]{
			\label{influenceofnneighbors:d}
			\includegraphics[width=0.46\linewidth]{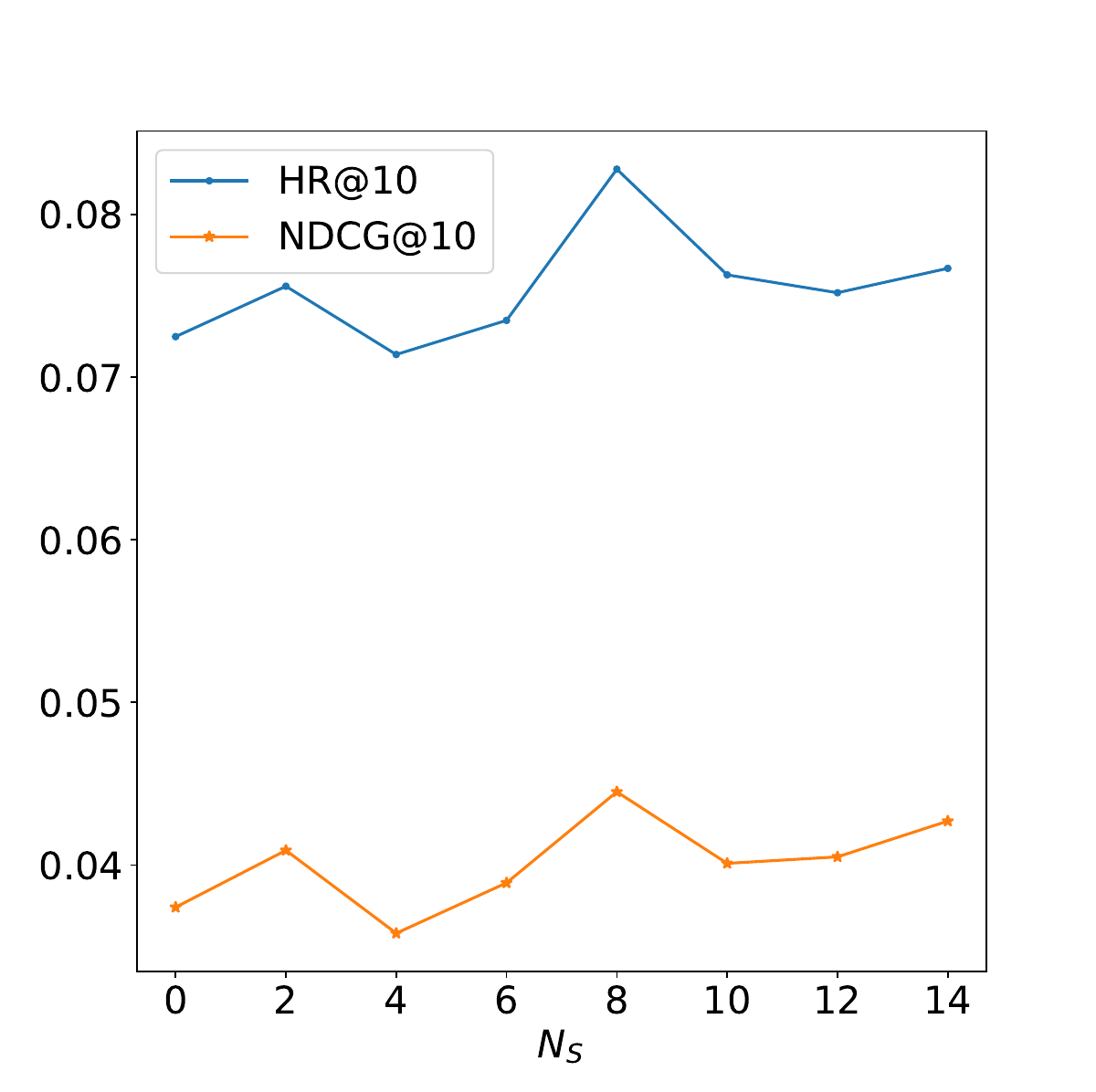}}
		\subfigure[PHP-dependency]{
			\label{influenceofnneighbors:a}
			\includegraphics[width=0.46\linewidth]{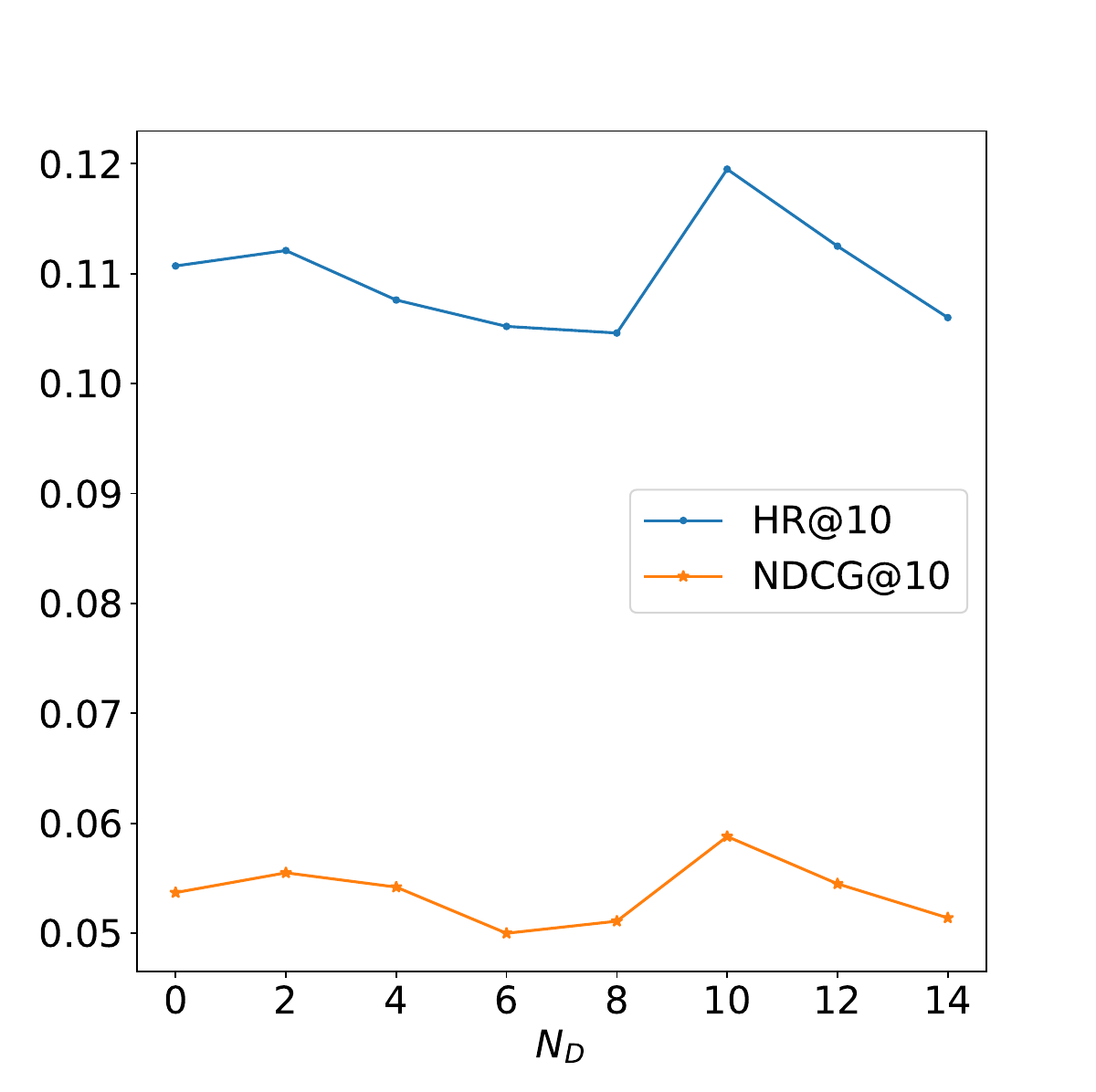}}
		\subfigure[Ruby-dependency]{
			\label{influenceofnneighbors:b}
			\includegraphics[width=0.46\linewidth]{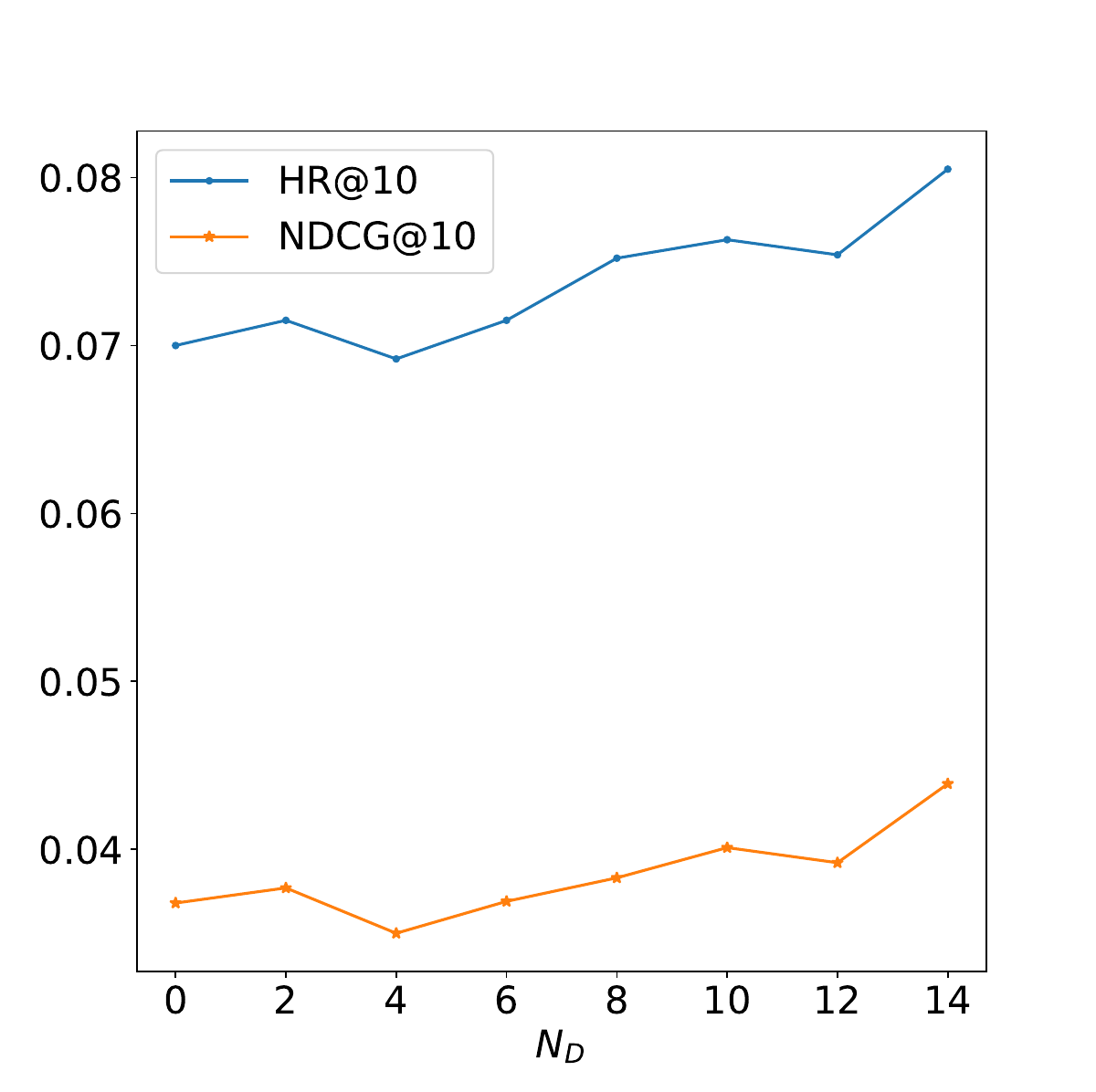}}
		\caption{Impact of neighborhood sample size.}
		\label{influenceofnneighbors}
	\end{figure}

    Due to the heterogeneity of both social and dependency networks, we utilize the sampling technique proposed in \cite{hamilton2017inductive} to ensure the training efficiency of the two graph attention networks. $|N_S|$ ($|N_D|$) neighbors are sampled for the first layer of the graph attention network for social network (dependency network) and the neighborhood sample size of each layer is half of the previous layer. In the analysis, we measure the recommendation performance using $HR@10$ and $NDCG@10$ under different neighborhood sample sizes as shown in Figure \ref{influenceofnneighbors}.
	
	From Figure \ref{influenceofnneighbors}, it can be found that neighborhood sample size for both graph attention networks for social and dependency networks has an impact on the recommendation performance, and too large or too small neighborhood sample size will both decrease the performance. The optimal neighbor sample sizes $N_S$ and $N_D$ for PHP are both 10, while they are 8 and 14 for Ruby, respectively.
	
    \subsubsection{Session length} \label{session-length}

\begin{figure}[H]
		\centering
		\subfigure[PHP]{
			\label{influenceoflifespan:a}
			\includegraphics[width=0.46\linewidth]{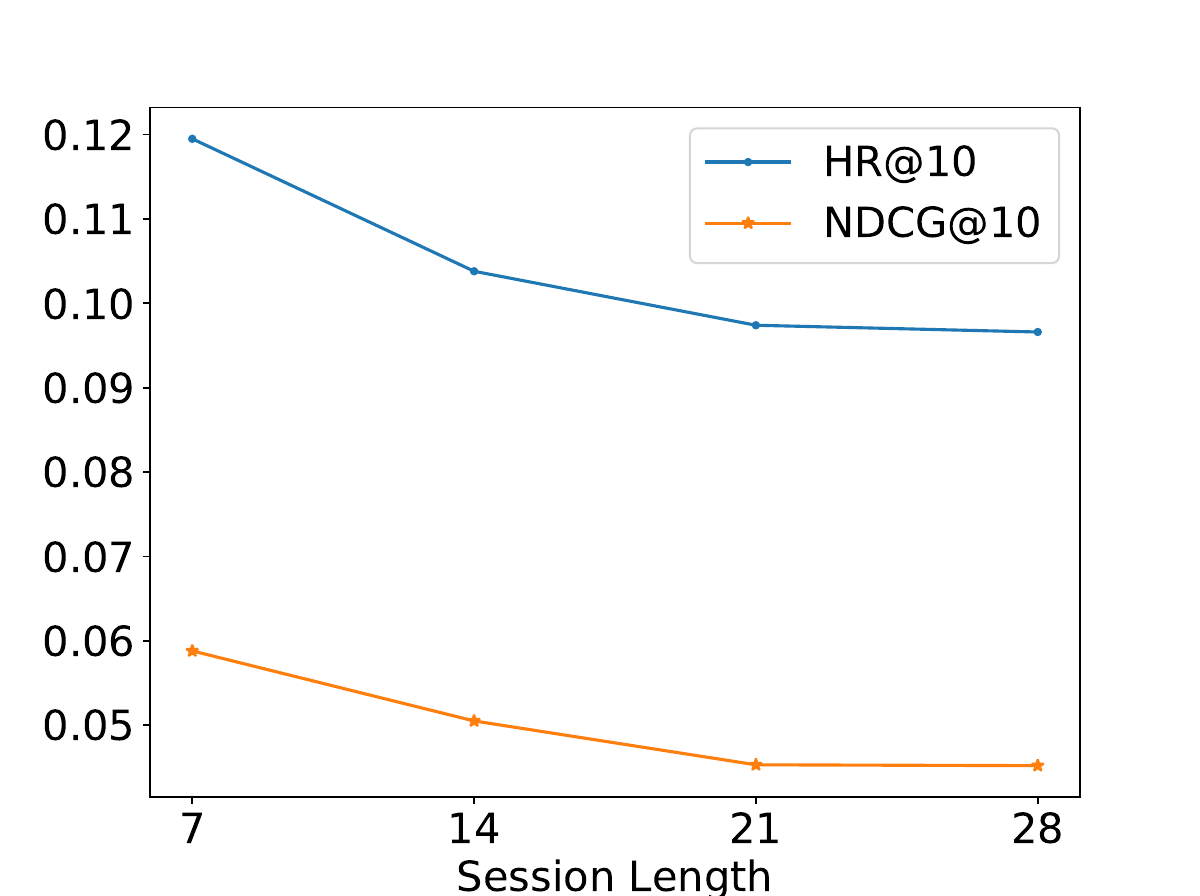}}
		\subfigure[Ruby]{
			\label{influenceoflifespan:b}
			\includegraphics[width=0.46\linewidth]{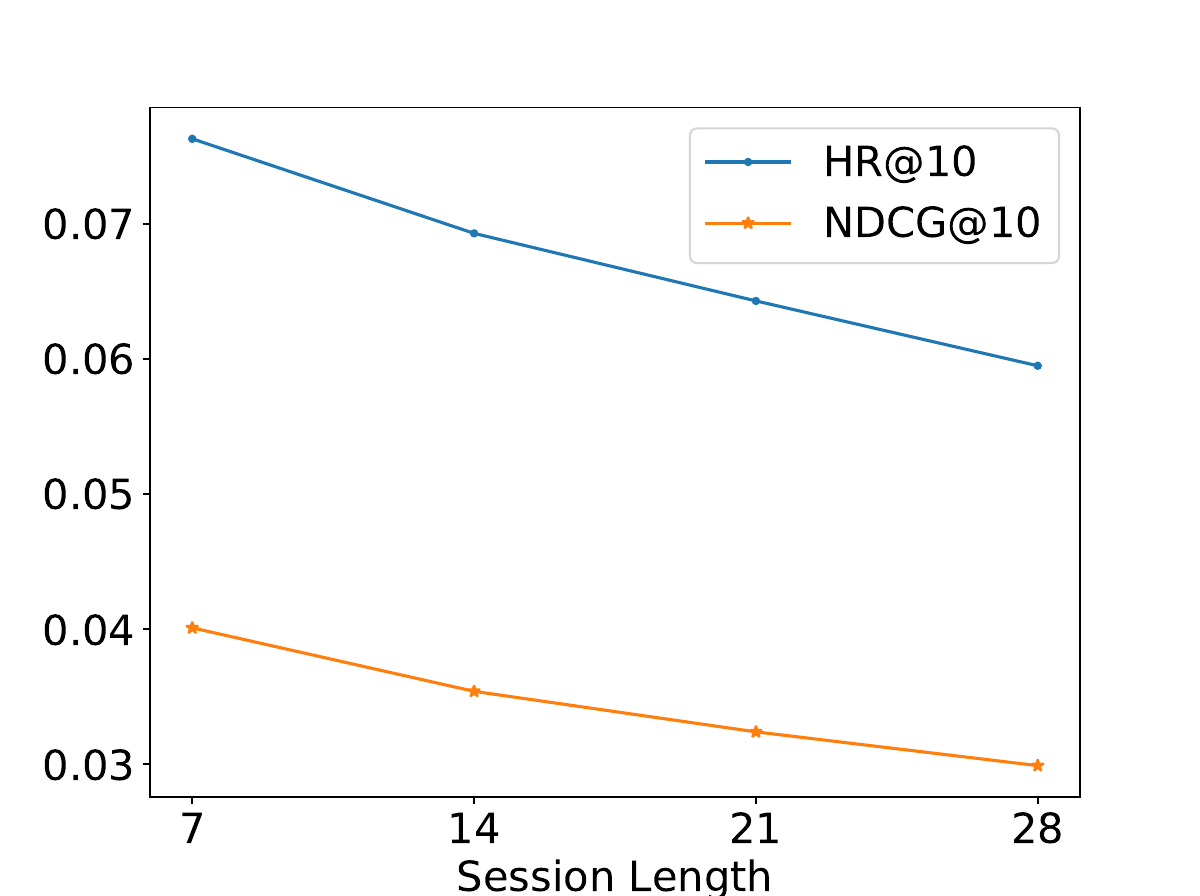}}
		\subfigure[JavaScript]{
			\label{influenceoflifespan:c}
			\includegraphics[width=0.46\linewidth]{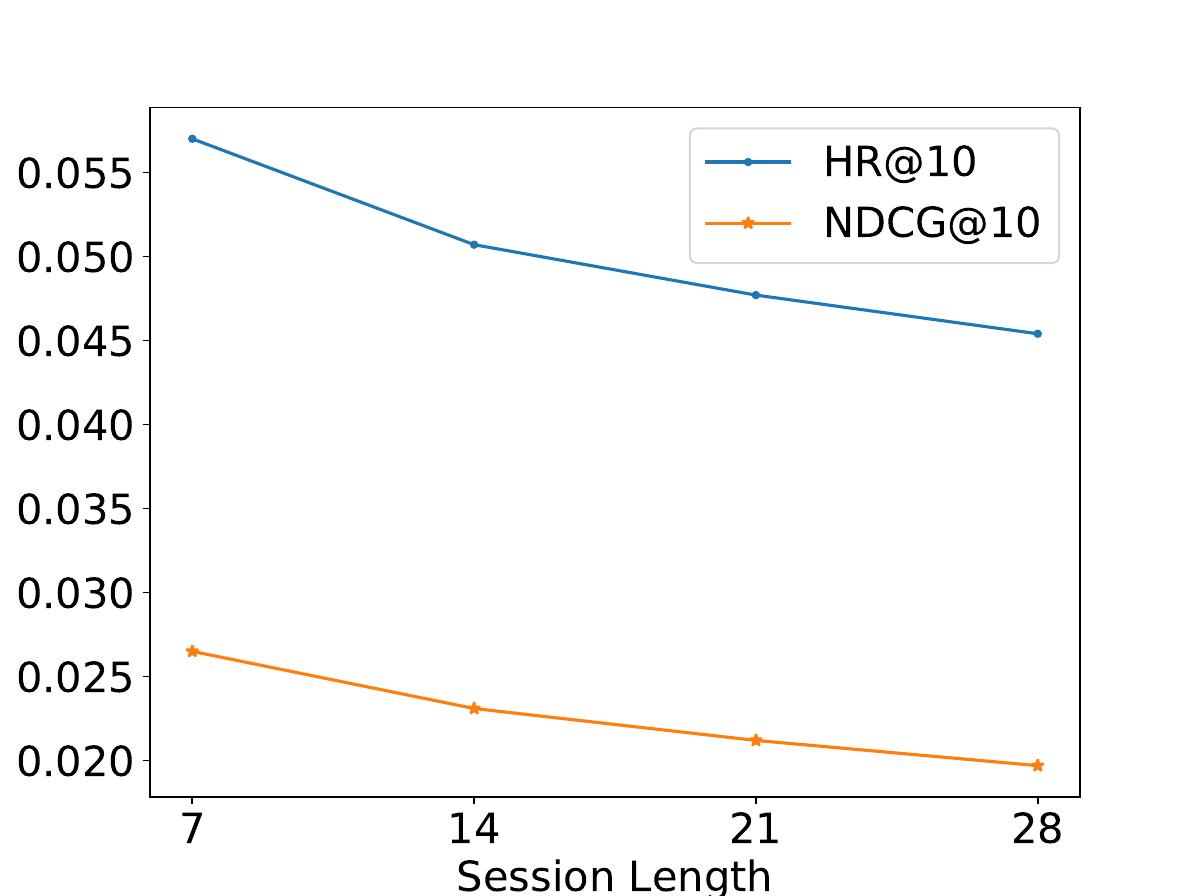}}
		\caption{Impact of session length.}
		\label{influenceoflifespan}
	\end{figure}

    Developers usually focus on a specific technical field during a certain length of time period, which is modeled as a session. Session length determines the granularity of dynamic interest modeling. In this analysis, we demonstrate the impact of session length by comparing the performance of SSDRec with different session lengths. From the result shown in Figure \ref{influenceoflifespan}, it can be observed that the performance decreases with the increase of session length and finally lies in a steady state. The reason is that longer session length means more coarse-grained granularity of dynamic interest. In an extreme setting where the whole interactions of a developer are segmented into one session, the RNN component of SSDRec captures the long-term static interest of the developer instead of short-term dynamic interest.
	
	\subsection{Attention Visualization} \label{Visualization}
	
    Section \ref{session-length} has demonstrated session can capture fine-grained interests. In this section, we further verify the hypothesis that developers' interests are relatively stable within a session but evolve across sessions by visualizing the attention weights of our model from the perspectives of both the overall development community and individual developer.
	
	\subsubsection{The overall development community}
	
\begin{figure}[H]
		\centering
		\subfigure[PHP]{
			\label{variance:a}
			\includegraphics[width=0.46\linewidth]{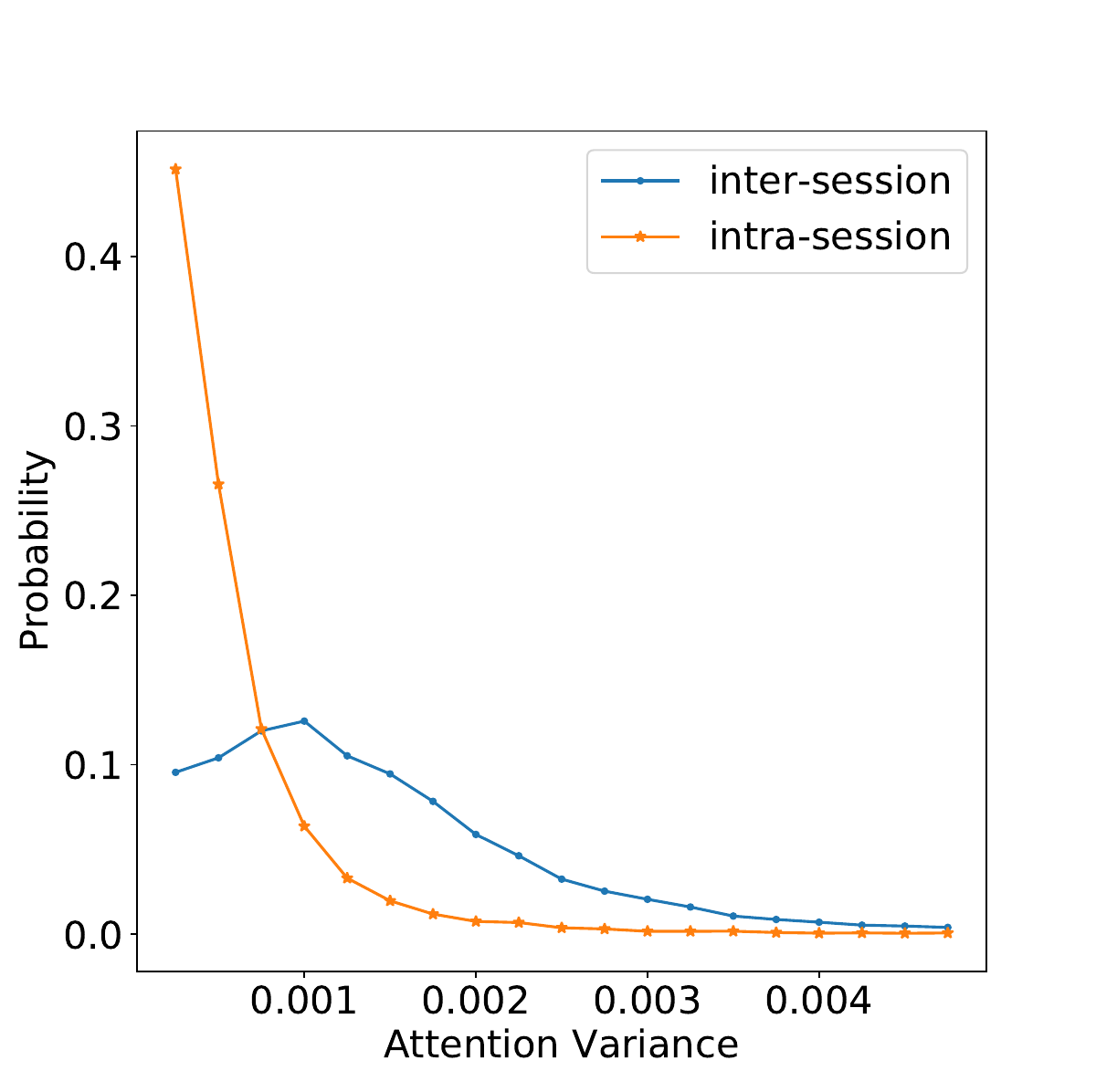}}
		\subfigure[Ruby]{
			\label{variance:b}
			\includegraphics[width=0.46\linewidth]{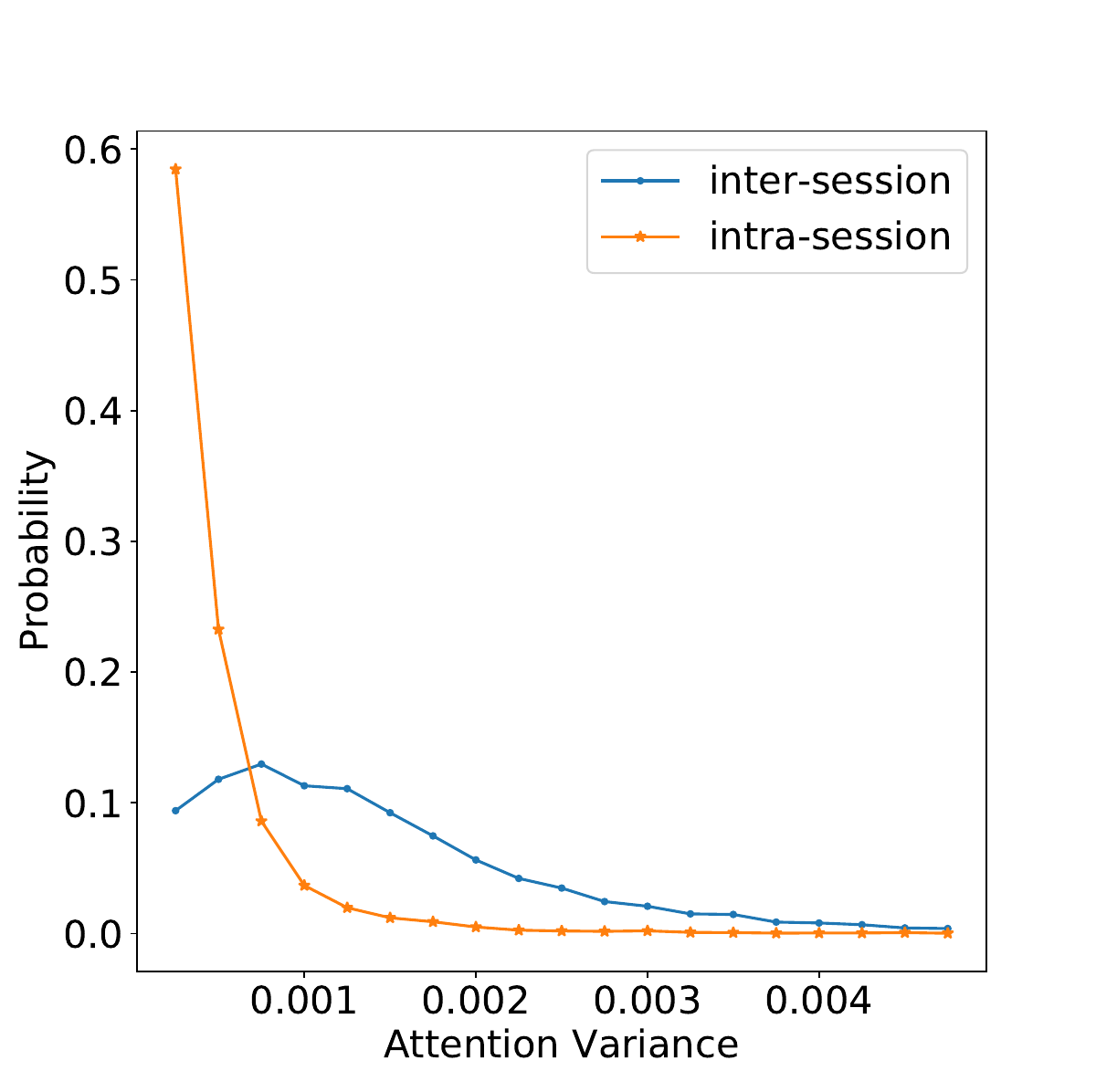}}
		\subfigure[JavaScript]{
			\label{variance:c}
			\includegraphics[width=0.46\linewidth]{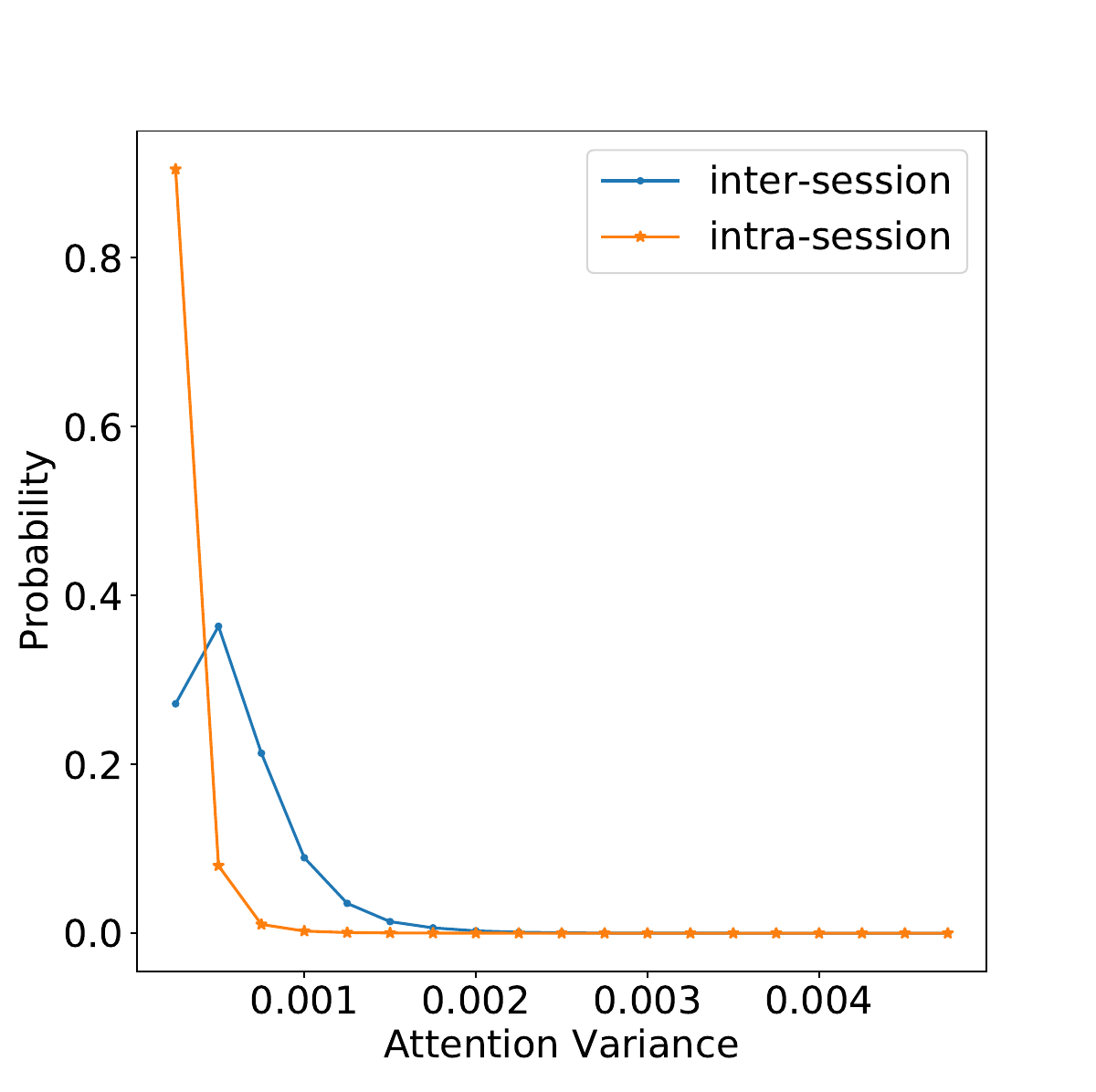}}
		\caption{Attention variance distributions of inter-session and intra-session}
		\label{variance}
	\end{figure}

    The attention weights between each pair of nodes of the social network across all sessions are grouped and the variance is then calculated as the inter-session attention variance. Similarly, the intra-session attention variance is calculated within each session. The distributions of both intra- and inter-session attention variance for PHP, Ruby and JavaScript are shown in Figure \ref{variance} from which we can observe that the attention weights across sessions vary largely than that within sessions. This confirms that developers' interests are relatively stable within a session but evolve across sessions.
	
	\subsubsection{Individual developer}
	
	In this section, we conduct a case study of an individual developer $u1039$ to show his/her behaviors within and across sessions. This developer has 8 test sessions and at least 5 friends in PHP community and the visualization results are shown in Figure \ref{user_invididual}. From the results, the following observations can be found:
	
	\begin{figure}[H]
		\centering
		\subfigure[intra-session attention]{
			\label{intra_sessions}
			\includegraphics[width=0.46\linewidth]{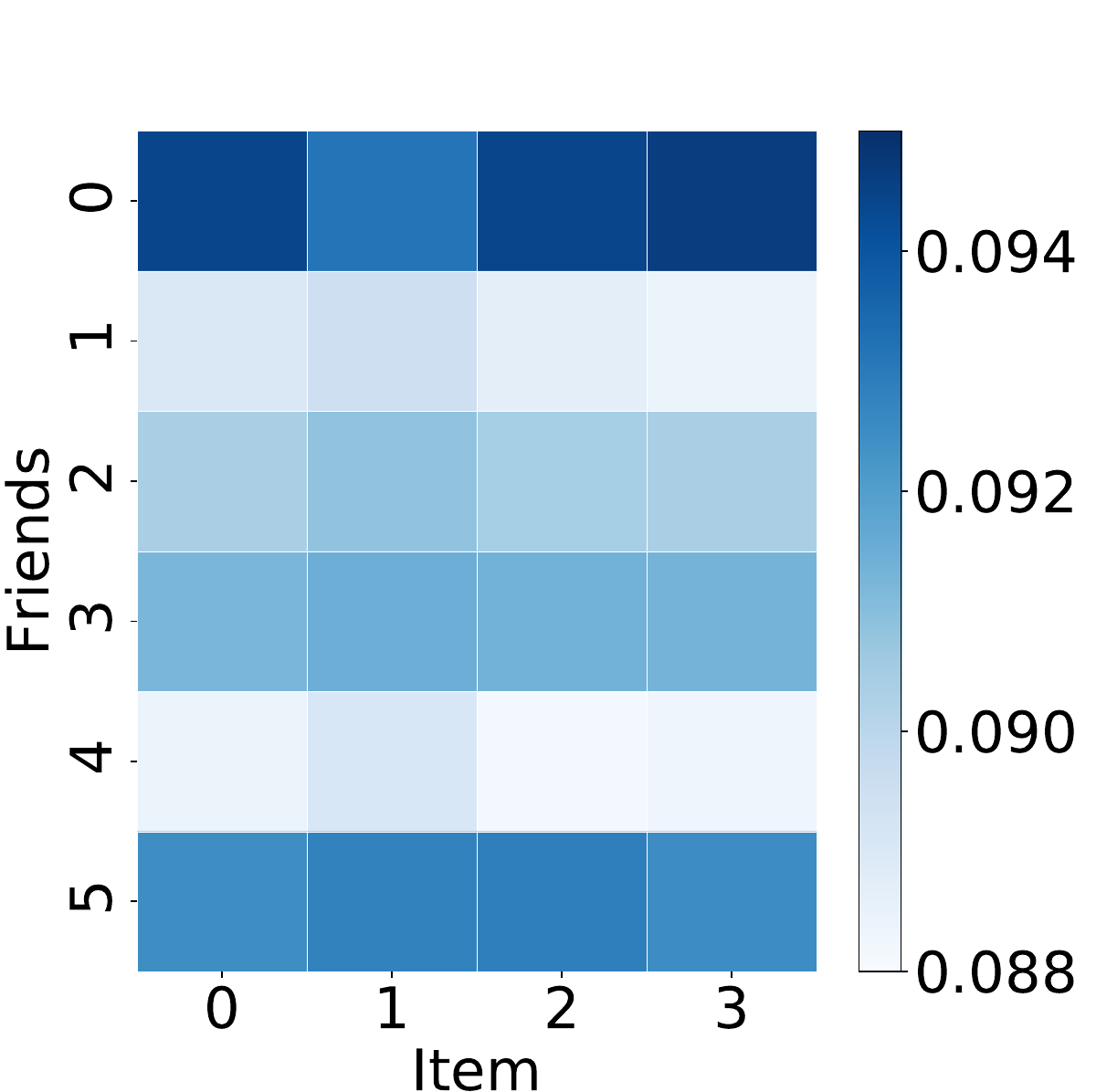}}
		\subfigure[inter-session attention]{
			\label{inter_sessions}
			\includegraphics[width=0.46\linewidth]{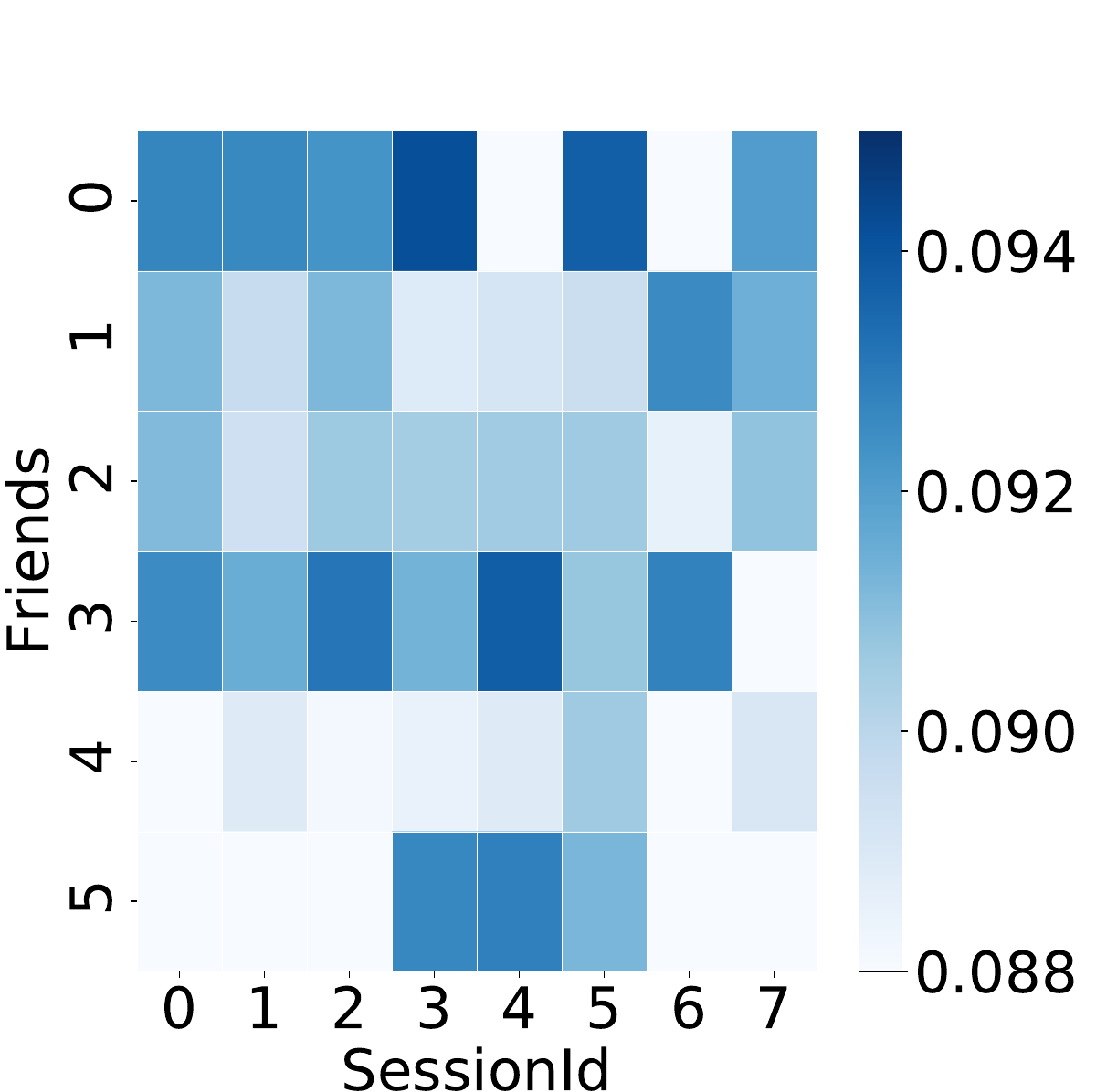}}
		\caption{Intra- and inter-session attention visualization for developer $u1039$.}
		\label{user_invididual}
	\end{figure}
	
    (1) The developer's interest is generally stable within the session and his/her technical choices are mainly influenced by friends 0 and 5.

    (2) The developer's interest evolves across sessions. During the first three sessions, i.e., session 0, 1 and 2, the developer's technical choices are mainly influenced by friends 0 and 3 while then in the three sessions that follow friend 5 begins to influence him/her. Maybe the developer begins to pay attention to a new technical field in session 3. Finally, the influence of friend 5 disappears in the last two sessions. The developer may leave this technical field after some investigation.

	\section{Conclusions}\label{CONCLUSIONS}

    In this article, we study the software recommendation problem and propose the Session-based Social and Dependency-aware software Recommendation model SSDRec to model the dynamic interests of developers with both social influence and dependency constraints in a unified framework. Especially, an RNN is employed to model the short-term dynamic interests and two GATs are utilized to capture social influence from friends and dependency constraints from dependent software packages, respectively. The experiments on real world datasets verify the effectiveness of all three components of our model. In the future, we will consider higher order relations in both social and dependency networks.
    
    \section*{CRediT authorship contribution statement}
    \textbf{Dengcheng Yan}: Conceptualization, Methodology, Writing - Original Draft. \textbf{Tianyi Tang}: Software, Validation, Data Curation, Writing - Original Draft. \textbf{Wenxin Xie}: Software, Validation.  \textbf{Yiwen Zhang}: Supervision, Writing - review \& editing. \textbf{Qiang He}: Writing - review \& editing.

    \section*{Declaration of Competing Interest}

    The authors declare that they have no known competing financial interests or personal relationships that could have appeared to influence the work reported in this paper.

    \section*{Acknowledgement}

    This work is supported by the National Key Research and Development Program of China (No. 2019YFB1704101), the National Natural Science Foundation of China (Grant No. 61872002, U1936220) and the University Natural Science Research Project of Anhui Province (Grant No. KJ2019A0037).

\bibliographystyle{unsrt}

\bibliography{SSDRec}

\end{sloppypar}
\end{document}